\documentclass[ASNA,twocolumn]{USG} %
\usepackage{anyfontsize} %

\usepackage{graphicx}   %
\usepackage{colortbl}   %
\usepackage{xcolor}     %
\usepackage{subfig}     %
\usepackage{algorithm}  %
\usepackage{algorithmicx}  %
\usepackage{algpseudocodex}%
\usepackage{tikz} %
\usepackage{fontawesome5}
\usepackage{siunitx}
\usepackage{microtype}

\usepackage{steinmetz}
\graphicspath{{./}}

\articletype{ORIGINAL ARTICLE}%

\received{\today}
\revised{00}
\accepted{00}
\journal{arXiv}
\volume{0}
\copyyear{2024}
\startpage{1}
\articledoi{10.1002/0000}

\begin{document}
\title{Evaluating Rust for Sparse Matrix Kernels in Scientific Computing}
\transtitle{Evaluating Rust for Sparse Matrix Kernels in Scientific Computing}
\subtranstitle{Evaluating Rust for Sparse Matrix Kernels in Scientific Computing}
\author[1]{Luca Lombardo}[https://orcid.org/0009-0001-1456-4478]%
\author[2]{Fabio Durastante}[https://orcid.org/0000-0002-1412-8289]

\authormark{Lombardo \textsc{et al.}}
\titlemark{Evaluating Rust for Sparse Matrix Kernels in Scientific Computing}

\address[1]{\orgdiv{Department of Computer Science}, \orgname{University of Pisa}, \orgaddress{\street{Largo Bruno Pontecorvo, 3}, \city{Pisa}, \postcode{56127}, \state{PI}, \country{Italy}}}

\address[2]{\orgdiv{Department of Mathematics}, \orgname{University of Pisa}, \orgaddress{\street{Largo Bruno Pontecorvo, 5}, \city{Pisa}, \postcode{56127}, \state{PI}, \country{Italy}}}

\corres{Fabio Durastante  (\email{fabio.durastante@unipi.it})}%


\transkeywords{Sparse matrix computation | Rust | Scientific Computing | Lanczos method}

\abstract[ABSTRACT]{Sparse matrix kernels form the computational backbone of scientific computing, traditionally relying on C/C++ and Fortran implementations that prioritize performance over memory safety. This work evaluates Rust as a systems-level alternative for sparse linear algebra by implementing and benchmarking three core workloads: sparse matrix-vector multiplication (SpMV), Lanczos-based Krylov methods, and matrix-exponential evaluation. We compare native Rust code against established baselines (Intel oneMKL, Eigen, PETSc, and PSBLAS) across a suite of representative matrices. Our results show that Rust's sparse kernels achieve performance comparable to Eigen and PSBLAS, tracking the state-of-the-art for CSC formats, while trailing PETSc's advanced blocked CSR optimizations. By analyzing compile-time monomorphization, SIMD vectorization, and FFI boundaries, we assess the practical impact of Rust's safety model and ecosystem readiness. The study provides concrete, evidence-based guidance for modernizing high-performance numerical software stacks.}

\transabstract[transABSTRACT]{Sparse matrix kernels form the computational backbone of scientific computing, traditionally relying on C/C++ and Fortran implementations that prioritize performance over memory safety. This work evaluates Rust as a systems-level alternative for sparse linear algebra by implementing and benchmarking three core workloads: sparse matrix-vector multiplication (SpMV), Lanczos-based Krylov methods, and matrix-exponential evaluation. We compare native Rust code against established baselines (Intel oneMKL, Eigen, PETSc, and PSBLAS) across a suite of representative matrices. Our results show that Rust's sparse kernels achieve performance comparable to Eigen and PSBLAS, tracking the state-of-the-art for CSC formats, while trailing PETSc's advanced blocked CSR optimizations. By analyzing compile-time monomorphization, SIMD vectorization, and FFI boundaries, we assess the practical impact of Rust's safety model and ecosystem readiness. The study provides concrete, evidence-based guidance for modernizing high-performance numerical software stacks.}

\maketitle

\section{Introduction}\label{sec:introduction}

Sparse matrix computations form the computational backbone of many scientific
and engineering applications~\cite{10.1098/rsta.2019.0053,1454783}, from solving partial differential equations~\cite{MR4993393} to network
analysis~\cite{MR4166335}. Traditionally implemented in C/C++ and Fortran, these computations demand
high performance and a coding framework which can be used with a contained effort
from domain scientists and engineers. The principal aim of this work is to evaluate
Rust, a modern  systems programming language emphasizing memory safety
without garbage collection, as a viable alternative for high-performance
sparse matrix operations both on the performance and usability fronts.

Recent publications indicate a growing interest in Rust for scientific and high-performance computing.
The role of Rust, together with Julia, as a modern option for scientific workloads and ecosystem
maturation is discussed in~\cite{10599948}. In numerical scientific computing, a Rust implementation of
a functional tensor-train library has been proposed for high-dimensional integration and PDE-oriented
workflows~\cite{10.1007/978-3-031-56208-2_22}, and Rust tooling for differential-equation solving is
documented in~\cite{Robinson2026}. At the library-ecosystem level, a GraphBLAS-oriented effort for Rust
is reported in~\cite{erbase}. From the formal-methods perspective, a refinement methodology for
distributed Rust programs is presented in~\cite{10.1145/3763119}, showing how Rust's ownership discipline
can support verification-oriented development. Together, these works motivate a deeper evaluation of Rust
in sparse linear algebra, where performance, safety, and software quality are all first-order concerns.

We present a comprehensive analysis of Rust's suitability for sparse linear algebra
by implementing and benchmarking key computational kernels: sparse matrix-vector
multiplication (SpMV), Lanczos methods as an orthogonalization technique for
the construction of Krylov subspaces, and matrix function evaluation techniques
based upon it. Our evaluation focuses on three critical dimensions: computational
performance relative to established C/C++ and Fortran implementations, %
the safety guarantees provided by Rust's ownership model and type system, and
the maturity of the Rust ecosystem for scientific computing.

The standardization of matrix operations has been crucial in establishing the
computational foundations of scientific computing. The Basic Linear Algebra
Subprograms (BLAS) standard~\cite{BLAS1979}, first introduced in the 1970s,
provided a unified interface for fundamental linear algebra operations and
became the de facto standard across implementations. This standardization
effort enabled the development of higher-level libraries such as LAPACK~\cite{LAPACK95},
which built upon BLAS kernels to provide robust implementations of advanced
numerical algorithms. Recognizing the need for similar standardization in
sparse matrix computations, the Sparse BLAS standard~\cite{SPARSEBLAS,abdelfattah2024interfacesparselinearalgebra} emerged to
extend  BLAS functionality to sparse data structures.

The whole standardization effort has been instrumental in defining
consistent interfaces for linear algebra operations, allowing implementations
across different programming languages and platforms to maintain
compatibility and performance portability~\cite{PSBLAS2000,PETSC2019,Ginkgo,PSCToolkit}.
These standards remain central to the scientific computing ecosystem, and any
modern language seeking to support sparse linear algebra must provide efficient bindings
or implementations that are compatible with or comparable to these established standards.

Through systematic benchmarking and numerical experiments on representative test
cases, we assess whether Rust can deliver competitive performance while providing
stronger safety guarantees than traditional languages. Our findings contribute to
the ongoing discussion on modernizing the scientific computing software stack~\cite{abdelfattah2024interfacesparselinearalgebra} and
provide practical insights for researchers considering Rust for computationally
intensive applications. We stress also that benchmark results for dense linear algebra kernels~\cite{faer_docs} are already available and have demonstrated highly promising performance\footnote{See \url{https://faer.veganb.tw/benchmarks/peddie6146/}.}, optimizing sparse operations remains a critical necessity. To ensure the reproducibility of what we show, the benchmarks are available on the GitHub repository~\href{https://github.com/lukefleed/hpla-rs}{github.com/lukefleed/hpla-rs}.

The remainder of this manuscript is organized as follows.
Section~\ref{subsec:sparse_background} reviews the role of sparse-matrix computations in scientific computing and provides background on the Rust programming language.
Section~\ref{sec:sparse_matrix_kernels} presents the sparse matrix kernels considered in our analysis.
Section~\ref{sec:implementation} describes the implementation of these kernels in Rust.
Section~\ref{sec:experimental_setup} outlines the experimental setup and presents the benchmarking results.
Finally, Section~\ref{sec:conclusions} summarizes our conclusions and discusses possible directions for future research.
 
\section{Sparse Matrix Computations in Scientific  Applications}\label{subsec:sparse_background}

Sparse matrices arise naturally in many scientific and engineering domains where
the underlying problem structure leads to systems with predominantly zero entries.
In computational fluid dynamics, finite element and finite difference discretizations
of partial differential equations produce sparse systems whose size can reach millions
or billions of unknowns~\cite{MR4993393}. Network analysis applications, such as graph algorithms and
social network computations, represent connectivity patterns as sparse adjacency matrices~\cite{MR4166335}.
In machine learning, recommendation systems and text analysis generate sparse matrices
from user-item interactions and term-document relationships.
\begin{figure*}[htbp]
	\centering
	\subfloat[Discretized PDE]{\includegraphics[height=0.7\columnwidth]{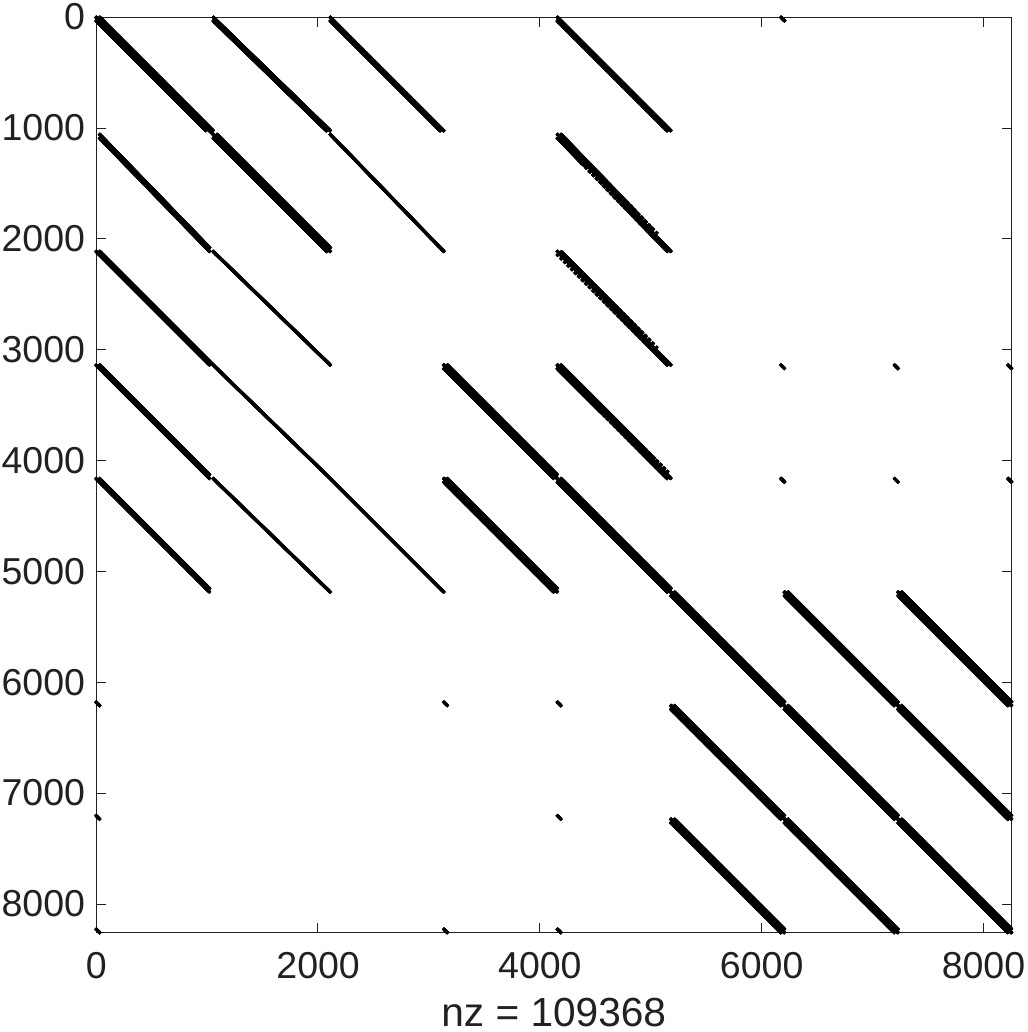}}\hfil
	\subfloat[Collaboration network]{\includegraphics[height=0.7\columnwidth]{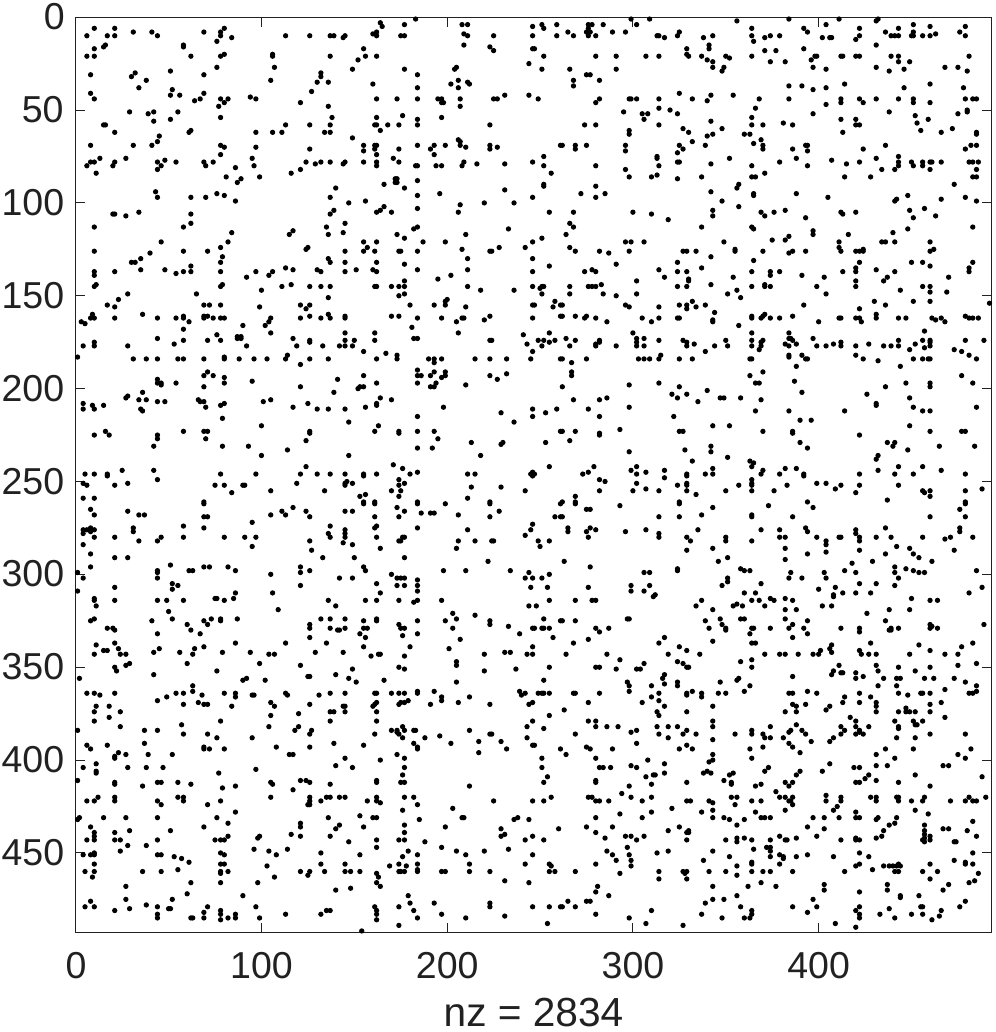}}
	\caption{Examples of sparse matrices with a high proportion of zero entries coming from discretized partial differential equations and a collaboration network; as depicted in the figure, these matrices display completely different levels of regularity in their non-zero distribution, ranging from highly structured patterns to highly irregular, random-like distributions.}
	\label{fig:sparse_matrix}
\end{figure*}

The key computational
challenge is that dense linear algebra algorithms, while theoretically applicable,
become prohibitively expensive in memory and computation time for large sparse problems. 
Consequently, specialized algorithms and data structures~\cite{FormatsOnGPUs,DesignPatternsSparseGPU,1454783} have been developed to exploit
sparsity patterns and store only non-zero entries. This exploitation of sparsity is
essential for solving real-world problems at scale, making efficient sparse matrix
computations a cornerstone of scientific and industrial computing. 

\subsection{The Rust Programming Language}\label{subsec:rust_language}

Rust~\cite{matsakis2014rust} is a systems programming language whose type system encodes ownership and borrowing as first-class constraints. Every value has exactly one owner, and assignment transfers ownership and invalidates the source, so the compiler identifies the precise point at which each allocation becomes unreachable and inserts deallocation there, requiring neither a garbage collector nor runtime reference counting. A static borrow checker proves that mutable references are exclusive and that all references remain valid for their entire scope, properties which in C and C++ the optimizer must approximate, or the programmer must annotate.

Rust's build system (\texttt{cargo}) normally compiles Rust dependencies from source as part of the same build graph. Within the Rust portion of a program, this gives the compiler visibility across library boundaries that in C and C++ are often hidden by separate compilation, except when templates or link-time optimization expose them again. The compiler exploits this visibility through a multi-level optimization pipeline: its central intermediate representation, Mid-level Intermediate Representation (MIR), is a typed control-flow graph on which it performs dataflow analyses and optimizations while the code is still generic, so that an optimization applied once benefits every concrete instantiation produced during the subsequent monomorphization step. Monomorphization generates dedicated machine code for each type, preserving the same inlining and vectorization opportunities available to non-generic code. The code generation backend is LLVM, the same infrastructure that Clang uses for C and C++. When combined with target-specific compilation flags, this model allows the Rust implementation to specialize generic numerical code for the target microarchitecture.

In sparse numerical codes, these mechanisms meet requirements that already arise from the algorithms and their measurement. A solver or benchmark keeps matrix storage alive across repeated kernel calls, lends views to library backends, updates work vectors under a fixed aliasing discipline, and separates setup allocations from the timed recurrence. Ownership, borrowing, traits, const generics, and explicit \texttt{unsafe} blocks give these requirements a place in the program structure. Furthermore, the Rust ecosystem now features a mature native backend for dense linear algebra called \texttt{faer}~\cite{faer_docs}, which eliminates the reliance on legacy foreign libraries for dense routines. We leverage this structure and native capability below to organize the benchmark harness, integrate \texttt{faer} whenever dense linear algebra operations are required, and manage both foreign backends and our native sparse kernels.

\section{Sparse Matrix Kernels}\label{sec:sparse_matrix_kernels}

This section introduces the computational kernels that form the technical core of our study: sparse matrix-vector multiplication, Lanczos-based Krylov subspace construction, and matrix-function evaluation through projected problems. The goal is twofold. First, we define the numerical operations and algorithmic building blocks that must be implemented in Rust and in the baseline languages. Second, we establish a common mathematical and computational framework that supports the implementation choices in Section~\ref{sec:implementation} and the performance analysis in Section~\ref{sec:results}.

\subsection{Sparse Matrix-Vector Product}\label{subsec:spmv}

A fundamental operation in sparse linear algebra is the sparse matrix-vector
product (SpMV). Given a sparse matrix \( A \) and a dense vector \( \mathbf{x} \), the SpMV
computes the product \( \mathbf{y} = \alpha A\mathbf{x} + \beta \mathbf{y} \), where
\( \mathbf{y} \) is the output vector, and \( \alpha \) and \( \beta \) are scalars. This operation is
ubiquitous in iterative methods for solving linear systems and eigenvalue problems,
as well as in various applications such as graph algorithms and machine learning.
The efficiency of SpMV implementations is critical, as it often constitutes the
computational bottleneck in large-scale simulations. Various storage formats, such as
Compressed Sparse Row (CSR) (Figure~\ref{fig:csr_data_and_spmv}) and Compressed Sparse Column (CSC), are employed to optimize
memory access patterns and reduce storage requirements. In here, we focus on the CSR format
due to its widespread use and efficiency in row-wise operations.
\begin{figure}[htbp]
	\centering
\subfloat[Sparse matrix in CSR format]{\begin{tikzpicture}[
		scale=0.75, transform shape,
		font=\normalsize,
		cell/.style={draw, minimum size=6mm, anchor=center, fill=gray!10},
		Label/.style={font=\bfseries\ttfamily, anchor=east}
		]

		\foreach \r/\c/\val/\col in {
			0/0/10/red!20, 0/3/-2/red!20,
			1/0/3/blue!20, 1/1/9/blue!20,
			2/1/7/green!20, 2/2/8/green!20, 2/3/7/green!20,
			3/0/3/orange!20, 3/2/-5/orange!20
		} {
			\node[cell, fill=\col] at (\c*0.6, -\r*0.6) (m-\r-\c) {$\val$};
		}

		\draw[thick] (-0.3, 0.3) rectangle (2.1, -2.1);
		\node[above] at (0.9, 0.4) {Matrix $A$};

		\node[Label] at (3.6, -0.2) {val};
		\foreach \i/\val/\c in {0/10/red!20, 1/-2/red!20, 2/3/blue!20, 3/9/blue!20, 4/7/green!20, 5/8/green!20, 6/7/green!20, 7/3/orange!20, 8/-5/orange!20} {
			\node[cell, fill=\c, anchor=west] (v\i) at (3.8 + \i*0.6, -0.2) {$\val$};
		}

		\node[Label] at (3.6, -0.9) {col\_ind};
		\foreach \i/\val/\c in {0/0/red!20, 1/3/red!20, 2/0/blue!20, 3/1/blue!20, 4/1/green!20, 5/2/green!20, 6/3/green!20, 7/0/orange!20, 8/2/orange!20} {
			\node[cell, fill=\c, anchor=west] (c\i) at (3.8 + \i*0.6, -0.9) {$\val$};
		}

		\node[Label] at (3.6, -1.6) {row\_ptr};
		\foreach \i/\val/\c in {0/0/red!20, 1/2/blue!20, 2/4/green!20, 3/7/orange!20, 4/9/white} {
			\node[cell, fill=\c, anchor=west] (r\i) at (3.8 + \i*0.6, -1.6) {$\val$};
		}
\end{tikzpicture}}

\subfloat[Pseudocode for the CSR SpMV]{%
	\begin{minipage}{0.48\textwidth} %
		\begin{algorithmic}[1]
			\Function{SpMV\_CSR}{$A, x, y, \alpha, \beta$}%
			\For{$i \gets 0$ \textbf{to} $n\_rows - 1$}%
			\State sum $\gets 0$%
			\For{$k \gets $\texttt{row\_ptr}$[i]$ \textbf{to} \texttt{row\_ptr}$[i+1] - 1$}%
			\State $j \gets $\texttt{col\_ind}$[k]$%
			\State sum $\gets \text{sum} + $\texttt{val}$[k] \times x[j]$%
			\EndFor%
			\State $y[i] \gets \alpha \times \text{sum} + \beta \times y[i]$%
			\EndFor%
			\EndFunction%
		\end{algorithmic}
	\end{minipage}%
}
	\caption{CSR data format and pseudocode for the CSR SpMV.}\label{fig:csr_data_and_spmv}
\end{figure}

\subsection{Krylov Subspace Methods: Lanczos Algorithm}\label{subsec:lanczos}

Krylov subspace methods are a class of iterative algorithms widely used for solving large
sparse linear systems~\cite{MR1990645} and eigenvalue problems~\cite{MR3396212}. These methods construct a sequence of
subspaces, known as Krylov subspaces, which are spanned by the successive applications
of the matrix \( A \) to an initial vector \( \mathbf{v}_1 \). The \( m \)-th Krylov subspace is defined as
$
\mathcal{K}_m(A, \mathbf{v}_1) = \text{span}\{\mathbf{v}_1, A\mathbf{v}_1, A^2\mathbf{v}_1, \ldots, A^{m-1}\mathbf{v}_1\}
$.
The Lanczos algorithm~\cite[\S 6.6]{MR1990645} is a prominent Krylov subspace method specifically designed for
symmetric matrices. It iteratively builds an orthonormal basis for the Krylov subspace
while simultaneously constructing a tridiagonal matrix \( T_m \) that approximates \( A \)
in the subspace. The matrix \( T_m \) can then be used to solve different problems, such as finding
approximate eigenvalues and eigenvectors of \( A \), approximating the solution of a linear system---in which
case the $T_m$ is only implicitly constructed and the orthogonalization is used inside the
Conjugate Gradient method---or evaluating matrix functions. We summarize the Lanczos algorithm
in Figure~\ref{alg:lanczos}, and illustrate the construction of the tridiagonal matrix \( T_m \)
in Figure~\ref{fig:lanczos-decomposition}.
\begin{figure}[htbp]
	\centering
	\subfloat[Lanczos Algorithm~\label{alg:lanczos}]{%
	\begin{minipage}{0.48\textwidth} 
		\begin{algorithmic}[1]
			\Require{Symmetric $A$, vector $\mathbf{v}_{1} \neq \mathbf{0}$, dimension $m$}
			\Ensure{$T_m = \operatorname{tridiag}(\boldsymbol{\beta},\boldsymbol{\alpha},\boldsymbol{\beta})$, orthogonal basis $V_m = [\mathbf{v}_1, \dots, \mathbf{v}_m]$}
			\State $\beta_1 \gets \| \mathbf{v}_1\|_2$, $\mathbf{v}_0 \gets \mathbf{0}$
			\State $\mathbf{v}_1 \gets \mathbf{v}_{1} / \beta_1$
			\For{$j = 1, 2, \dots, m-1$}
			\State $\mathbf{w}_j \gets A \mathbf{v}_j - \beta_j \mathbf{v}_{j-1}$\label{line:spmv}
			\State $\alpha_j \gets \langle \mathbf{w}_j, \mathbf{v}_j \rangle$\label{line:dot}
			\State $\mathbf{w}_j \gets \mathbf{w}_j - \alpha_j \mathbf{v}_j$\label{line:axpy}
			\State $\beta_{j+1} \gets \|\mathbf{w}_j\|_2$\label{line:nrm2}
			\If{$\beta_{j+1} \approx 0$}
			\State \textbf{break}
			\EndIf
			\State $\mathbf{v}_{j+1} \gets \mathbf{w}_j / \beta_{j+1}$\label{line:scal}
			\EndFor
		\end{algorithmic}
	\end{minipage}%
	}
	
	\subfloat[Tridiagonal matrix \( T_m \) generated by the Lanczos algorithm\label{fig:lanczos-decomposition}]{
		\centering
		\begin{tikzpicture}[
			scale=0.7, transform shape,
			font=\normalfont,
			border/.style={draw, thick},
			fill_gray/.style={fill=gray!20}
			]
			
			\node at (3.2, 0.6) {Lanczos Algorithm};

			\draw[border, fill_gray] (0, 0) rectangle (2, -2);
			\node at (1, -1) {$A$};

			\draw[border, fill_gray] (2.5, 0) rectangle (3.5, -2);
			\node at (3, -1) {$V_m$};

			\node at (4, -1) {$\approx$};

			\draw[border, fill_gray] (4.5, 0) rectangle (5.5, -2);
			\node at (5, -1) {$V_m$};

			\draw[border, fill_gray] (6, 0) rectangle (7, -1);
			\node at (6.5, -0.5) {$T_m$};
			
		\end{tikzpicture}\hspace{1em}
		
		\begin{tikzpicture}[
			scale=0.8, transform shape,
			font=\normalfont,
			cell/.style={draw, minimum size=6mm, anchor=center, fill=gray!10},
			Label/.style={font=\bfseries\ttfamily, anchor=east}
			]

			\foreach \r/\c/\val/\col in {
				0/0/$\alpha_1$/red!20, 0/1/$\beta_2$/red!20,
				1/0/$\beta_2$/blue!20, 1/1/$\alpha_2$/blue!20, 1/2/$\beta_3$/blue!20,
				2/1/$\beta_3$/green!20, 2/2/$\alpha_3$/green!20, 2/3/$\beta_4$/green!20,
				3/2/$\beta_4$/orange!20, 3/3/$\alpha_4$/orange!20
			} {
				\node[cell, fill=\col] at (\c*0.6, -\r*0.6) (t-\r-\c) {\val};
			}

			\draw[thick] (-0.3, 0.3) rectangle (2.7, -2.7);
			\node[above] at (1.2, 0.4) {$T_m$};
			
		\end{tikzpicture}}
	\caption{Lanczos algorithm for constructing a tridiagonal matrix \( T \) and
		an orthonormal basis \( V_m \) for the Krylov subspace \( \mathcal{K}_m(A, \mathbf{v}_1) \).\label{fig:lanczos}}
\end{figure}
We observe that the operation \( A \mathbf{v}_j \) in line 4 of the algorithm is a sparse matrix-vector product,
which is the computational kernel discussed in Section~\ref{subsec:spmv}, and the other operations we
need are the BLAS level 1 operations for vector updates (at line \ref{line:axpy}), inner product/norm (at lines \ref{line:dot} and \ref{line:nrm2}), and
vector scaling (at line \ref{line:scal}). This implies that we also need an efficient implementation
of these kernels available in Rust to have a performant implementation of the Lanczos algorithm, and we rely on the faer library on rust side~\cite{faer_docs}.
For the term of comparison with different language implementations, we can call the relevant BLAS routines
and link against the most performant BLAS implementation available on the target system, e.g., the \texttt{Intel MKL} or the \texttt{OpenBlas} libraries.

\subsection{Matrix Function Evaluation}\label{subsec:matrix_functions}

Matrix functions, such as the matrix exponential, logarithm, and square root, are of significant
interest in various applications, including network analysis, quantum mechanics, control theory,
and the construction of exponential integrators for the solution of initial value problems. For
a generic matrix \( A \), we can always define a matrix function \( f(A) \) via the Jordan canonical
form~\cite[Definition~1.2, p.3]{MR2396439} or, in a typically more computationally relevant way, via the Cauchy integral formula~\cite[Definition~1.1, p.8]{MR2396439}:
\[
f(A) = \frac{1}{2\pi i} \oint_{\Gamma} f(z) (zI - A)^{-1} dz,
\]
where \( \Gamma \) is a closed contour in the complex plane enclosing the spectrum of \( A \).
As far as sparse matrices are concerned, the direct computation of \( f(A) \) is often infeasible due to
the size and sparsity of \( A \). Instead, we typically seek to compute the action
of the matrix function on a vector \( \mathbf{b} \), denoted as \( f(A)\mathbf{b} \).
This approach avoids the explicit formation of \( f(A) \) on the large sparse matrix \( A \), and
uses Krylov subspace methods to approximate the result~\cite{MR1149094}. To simplify the exposition,
we focus here on the case of a symmetric matrix \( A \) and a function \( f \) that is analytic
on an interval $[a,b] \subseteq \mathbb{R}$ containing the spectrum of \( A \). The approximation of \( f(A)\mathbf{b} \).
can be achieved by projecting the problem onto the Krylov subspace \( \mathcal{K}_m(A, \mathbf{b}) \)
using the Lanczos algorithm described in Section~\ref{subsec:lanczos}. The approximation is given by
\begin{equation}\label{eq:lanczos_mat_fun}
	f(A)\mathbf{b} \approx \beta_1 V_m f(T_m) \mathbf{e}_1,
\end{equation}
where \( V_m \) is the matrix whose columns are the orthonormal basis vectors generated
by the Lanczos algorithm, \( T_m \) is the tridiagonal matrix constructed during the process,
\( \beta_1 = \|\mathbf{b}\|_2 \), and \( \mathbf{e}_1 \) is the first canonical basis vector in \( \mathbb{R}^m \).
The computation of \( f(T_m) \) is feasible since \( m \) is typically much smaller than the size of \( A \), and various methods,
such as diagonalization or Pad\'{e} approximants, can be employed to evaluate \( f(T_m) \) efficiently~\cite{MR2396439}.

A critical aspect of this approximation is determining the dimension \( m \) required to achieve a desired
accuracy. Unlike solving linear systems where the residual \( \|\mathbf{b} - A\mathbf{x}_m\| \) is readily
available, the error in matrix function approximation is harder to estimate directly~\cite{MR1149094,ErrorBoundsLanczos2022}.

To provide a standard and widely adopted benchmark, we consider the computation of the matrix exponential. This matrix function is of practical importance, appearing in applications ranging from exponential integrators for ordinary differential equations~\cite{Expokit} to the analysis of complex networks~\cite{MR4166335}. At the same time, the conclusions drawn from this study are not specific to the exponential, since most of the following considerations apply equally to other matrix functions, such as the logarithm or the square root.
If we chose $f(z)=e^{t\,z}$, $t \in \mathbb{R}$, in~\eqref{eq:lanczos_mat_fun}, the analysis in~\cite{MR1149094} gives an explicit error representation from which one recovers a simple and effective \emph{a posteriori} estimate based on the small projected matrix $T_m$.
In particular, the leading term of the error expansion suggests the approximation
\begin{equation}\label{eq:saad_bound}
	\begin{split}
		\| \exp(tA) \mathbf{b} - \beta_1 V_m \exp(t T_m) \mathbf{e}_1 \|_2 & \approx \eta_m(t) \\ & =
		\beta_1\,|\beta_{m+1}|\,
		\left|\mathbf{e}_m^\top e^{tT_m}\mathbf{e}_1\right|.
	\end{split}
\end{equation}
This remains an approximation which is good in practice, and is adequate for the task
of running our numerical experiments and comparing the performance of different implementations,
since we are interested in the relative performances within a reasonable accuracy range, rather than in the absolute error values;
in general we refer back to~\cite{MR1149094,ErrorBoundsLanczos2022} for a more detailed discussion on the error representation and estimation for Lanczos-based matrix function approximations.

Before focusing on the small-scale computation of \( f(T_m) \), we need the following observation.
Let us suppose that storing $V_m$, which costs $n \times m$ the size of the used datatype,
is higher than the storage space we can commit to this task. In this case, we can use a
\emph{two-pass} Lanczos procedure, where in the first pass we only compute the
tridiagonal matrix \( T_m \), and in the second pass we reconstruct the basis vectors \( V_m \)
one at a time to compute the final result \( f(A)\mathbf{b} \). This approach reduces memory
requirements at the cost of additional computational effort, i.e., we double the number of
matrix-vector multiplications. Crucially, however, the re-computation of the basis vectors does
not require repeating the scalar products for orthogonalization; the coefficients $\alpha_j$ and
$\beta_j$ have already been computed and stored in the tridiagonal matrix $T_m$ during the first
pass and can be directly recovered. In many applications, these memory savings may end
up outweighing the extra computations; we summarize the two-pass Lanczos method for the matrix exponential in Figure~\ref{fig:two-pass-lanczos}.
\begin{figure}[htb]
	\begin{algorithmic}[1]
		\Require{Symmetric $A$, vector $\mathbf{b}$, time $t\in\mathbb{R}$, dimension $m$}
		\Ensure{$\mathbf{y}_m(t) \approx e^{tA}\mathbf{b}$ and error indicator $\eta_m(t)$}
		\State \textbf{First Pass}
		\State Run $m$ Lanczos steps with start vector $\mathbf{b}$ and store $\{\alpha_j\}_{j=1}^m$, $\{\beta_j\}_{j=1}^{m+1}$
		\State Form $T_m=\operatorname{tridiag}(\beta_2,\ldots,\beta_m;\alpha_1,\ldots,\alpha_m)$
		\State Compute $\mathbf{g}=\beta_1 e^{tT_m}\mathbf{e}_1$, where $\mathbf{g}=[g_1,\ldots,g_m]^\top$
		\State Compute $\eta_m(t)=\beta_1\,|\beta_{m+1}|\,\left|\mathbf{e}_m^\top e^{tT_m}\mathbf{e}_1\right|$
		\State \textbf{Second Pass}
		\State $\beta_1 \gets \| \mathbf{b}\|_2$, $\mathbf{v}_0 \gets \mathbf{0}$
		\State $\mathbf{v}_1 \gets \mathbf{b} / \beta_1$
		\State $\mathbf{y}_m \gets g_1 \mathbf{v}_1$
		\For{$j = 1, 2, \dots, m-1$}
		\State $\mathbf{w}_j \gets A \mathbf{v}_j - \beta_j \mathbf{v}_{j-1}$
		\State $\mathbf{w}_j \gets \mathbf{w}_j - \alpha_j \mathbf{v}_j$
		\State $\mathbf{v}_{j+1} \gets \mathbf{w}_j / \beta_{j+1}$
		\State $\mathbf{y}_m \gets \mathbf{y}_m + g_{j+1} \mathbf{v}_{j+1}$
		\EndFor
	\end{algorithmic}
	
	\caption{Two-pass Lanczos method for approximating $e^{tA}\mathbf{b}$ and evaluating the practical indicator $\eta_m(t)$.\label{fig:two-pass-lanczos}}
\end{figure}

Now, for the small-scale computation of \( f(T_m) \), since \( T_m \) is a symmetric tridiagonal
matrix, it can be efficiently diagonalized, allowing us to compute
\( f(T_m) \) via the spectral decomposition $T_m = Q \Lambda Q^\top$ giving $f(T_m) = Q f(\Lambda) Q^\top$,
for \( Q \) is the orthogonal matrix of eigenvectors and \( \Lambda \) is the diagonal matrix of eigenvalues.
The function \( f(\Lambda) \) is computed by applying \( f \) to each diagonal entry of \( \Lambda \).
This approach is computationally efficient for small \( m \) and leverages well-established
numerical linear algebra techniques. Again, for the term of comparison with different
language implementations we can fall-back to the LAPACK routine \texttt{xSTEV} to compute
the eigendecomposition of \( T_m \), and the \texttt{xGEMM} routine to compute
the products \( Q f(\Lambda) Q^\top \mathbf{e}_1 \).

\section{Implementation in Rust}\label{sec:implementation}

The experimental comparison follows this structure. A single source of matrix data feeds all backends, and a fixed boundary separates construction from repeated kernel execution. The Rust harness owns the sparse storage used to initialize every backend, lends borrowed views to libraries that accept external arrays, allocates matrix-dependent workspaces during setup, and enters each timed kernel with the relevant shape, layout, lifetime, and aliasing assumptions already fixed. The Rust code also supplies the native SpMV and Lanczos implementations. We first describe the data model and setup boundary, then turn to the Lanczos workspace, the specialized SpMV kernel, and the build model for foreign backends.

\subsection{Benchmark data model and ownership}\label{subsec:data_model_ownership}

To measure kernel execution rather than matrix parsing or format conversion, the benchmark harness parses each Matrix Market file once and materializes, from the same input, the compressed-row, compressed-column, and triplet representations required by the benchmarked implementations. Rust owns these buffers for the lifetime of the benchmark instance, so backends that accept external storage receive stable borrowed views. When the Lanczos experiments rescale the operator, the scaling is applied before backend setup and to every stored representation, so all backends are initialized with the same numerical matrix.

Once these representations have been fixed, each backend performs the setup required by its own interface. PETSc and Eigen borrow the Rust-owned compressed arrays, and MKL may construct an optimized internal representation during its inspection phase. PSBLAS creates its context and topology descriptor, builds an identity local-to-global map for the single-process run, inserts the sparse entries through \texttt{psb\_c\_dspins}, and finalizes an internal CSR or CSC matrix with \texttt{psb\_c\_dspasb\_opt}. Since these operations occur before timing starts, the measured loop contains backend execution on already constructed operands, while matrix loading, format construction, backend setup, inspection, and assembly remain outside the reported kernel cost.

\subsection{Kernel execution model}\label{subsec:kernel_execution_model}

Sparse kernels are commonly applied many times after the sparse operator and the working vectors have been constructed, so setup costs and execution costs must be separated. Format construction, sparse-handle creation, inspection, assembly, and harness-owned workspace allocation may scale with \(n\), \(\operatorname{nnz}\), or \(nm\), but they are not paid at every application of an already constructed operator. We place these operations before the timing loop, and the timed region starts only after the backend context, dense vectors, sparse objects, and harness-owned workspaces have been constructed. Work arrays allocated internally by a library execution routine remain part of that backend measurement.

Each timed iteration calls the backend execution routine on these pre-existing operands. In the SpMV benchmark this routine applies the sparse matrix to a fixed dense vector, while in the Lanczos benchmarks it also performs the vector updates, inner products, normalizations, projected tridiagonal computation, and final reconstruction required by the selected algorithm. The native Rust implementation follows the same boundary by allocating workspaces whose size depends on \(n\), \(\operatorname{nnz}\), or \(nm\) before timing starts and reusing them across iterations. The reported time therefore excludes matrix loading, format construction, and explicit backend setup, but includes any work that a backend performs inside its own execution call.

\subsection{Lanczos storage and data movement}\label{subsec:lanczos_storage}

The storage layout is determined by the information needed after the recurrence. The one-pass variant must retain the basis \(V_m\), so its dominant storage term is \(O(nm)\). The two-pass variant reconstructs the approximation without retaining \(V_m\), reducing the matrix-dependent storage to three length-\(n\) rolling vectors and the output vector, at the cost of a second sequence of sparse matrix-vector products. The recurrence coefficients and the projected vector \(\mathbf{g}_m=\exp(-T_m)\mathbf{e}_1\) require \(O(m)\) storage. Since all implementations evaluate the projected problem by diagonalizing \(T_m\), they also store the eigenvectors of the tridiagonal matrix, giving an \(O(m^2)\) term independent of the original matrix dimension. This term is bounded by the chosen Krylov dimension and does not change the distinction between the \(O(nm)\) one-pass basis storage and the \(O(n)\) two-pass recurrence storage.

The Rust implementation follows this layout by storing the basis, rolling vectors, scalar arrays, and output vector in a workspace that the driver splits into disjoint mutable borrows before entering the recurrence. The sparse operator and the dense vectors are passed as borrowed views, so the aliasing constraints of the update are expressed in the type system. During the recurrence, the previous vector, current vector, and residual work vector are rotated by swapping their matrix headers, so advancing one step changes the logical role of already allocated buffers without copying length-\(n\) vector contents.

The recurrence is also organized to reduce vector passes. After the sparse product has written \(A \mathbf{v}_j\) into the residual buffer, the subtraction of the previous-vector term and the accumulation of the Rayleigh quotient are fused in the common case \( \beta_{j-1} \neq 0 \). The subtraction of the current-vector term is then delegated to faer's dense kernel path~\cite{faer_docs}. On the last iteration, the driver stops after computing the residual norm, because no subsequent recurrence step consumes the normalized residual or the rotated buffers.

\subsection{Specialized CSR sparse matrix-vector product}\label{subsec:csr_spmv_kernel}

In faer's generic path, CSR sparse-dense multiplication is reduced to the CSC kernel by forming transposed borrowed views of the sparse matrix, dense input, and output. This view construction is zero-copy because the transposed views reuse the same index, value, and dense storage, allowing the library to share one compressed-column implementation across the two formats. The same abstraction, however, hides the row-wise CSR traversal needed for the sequential single-vector regime considered in the benchmarks. We therefore added a direct CSR path that traverses rows explicitly.\footnote{\url{https://codeberg.org/lukefleed/faer/src/branch/main/faer/src/sparse/linalg/csr_spmv.rs}} This path is one implementation contribution of this work. The specialization is deliberately narrow because it should affect the benchmarked kernel without changing the behavior of the generic sparse interface. It is selected only for \texttt{u32} indices, \texttt{f64} values, one right-hand side, one output column, and sequential execution. All other cases keep the generic transpose-through-CSC path. While this specialization demonstrates that Rust can emit competitive SIMD assembly, extending this kernel to support a wider array of datatypes (e.g., single-precision and complex numbers) and non-AVX-512 vector paths (e.g., AVX2 and ARM NEON) remains a requirement that can be implemented.

The public interface performs shape and type checks before entering the kernel. The specialized path then casts the checked storage once and dispatches on the remaining loop-invariant cases. The accumulation mode distinguishes replacement from addition, the scaling case distinguishes \(\alpha=1\) from \(\alpha\neq1\), and the dense-vector layout distinguishes unit-stride from strided access. Padded and unpadded CSR rows are also separated before the row loop. These choices are fixed for one kernel call, so the implementation encodes them in trait and const-generic parameters and lets Rust monomorphize the selected row-loop instance. The hot loop therefore sees only row bounds, column indices, numerical values, the input-vector pointer, and the output slot.

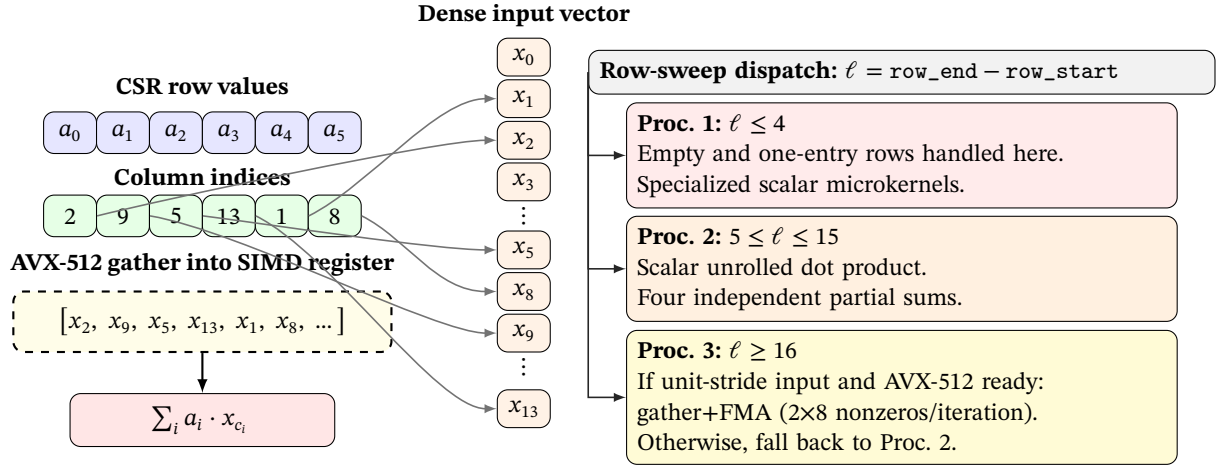
\begin{figure*}[hbtp]
	\centering
	\begin{tikzpicture}[
		font=\normalsize,
		>=latex,
		box/.style={draw, rounded corners, minimum height=0.55cm, minimum width=0.7cm, align=center},
		val/.style={box, fill=blue!10},
		idx/.style={box, fill=green!10},
		vec/.style={box, fill=orange!10, minimum height=0.5cm},
		acc/.style={box, fill=red!10},
		gather/.style={->, semithick, draw=black!55},
		dispatcharrow/.style={->, semithick, draw=black!85},
		]

		\node[val] (a0) at (0,0.1) {$a_0$};
		\node[val] (a1) at (0.7,0.1) {$a_1$};
		\node[val] (a2) at (1.4,0.1) {$a_2$};
		\node[val] (a3) at (2.1,0.1) {$a_3$};
		\node[val] (a4) at (2.8,0.1) {$a_4$};
		\node[val] (a5) at (3.5,0.1) {$a_5$};
		\node[above, font=\bfseries] at (1.75,0.5) {CSR row values};

		\node[idx] (c0) at (0,-1) {$2$};
		\node[idx] (c1) at (0.7,-1) {$9$};
		\node[idx] (c2) at (1.4,-1) {$5$};
		\node[idx] (c3) at (2.1,-1) {$13$};
		\node[idx] (c4) at (2.8,-1) {$1$};
		\node[idx] (c5) at (3.5,-1) {$8$};
		\node[above, font=\bfseries] at (1.75,-0.7) {Column indices};

		\node[
		draw, thick, dashed, rounded corners,
		minimum width=5.0cm, minimum height=0.8cm, fill=yellow!10
		] (simd) at (1.75,-2.4)
		{
			$\big[x_2,\;x_9,\;x_5,\;x_{13},\;x_1,\;x_8,\;\ldots\big]$
		};
		\node[above, font=\bfseries] at (1.75,-1.9) {AVX-512 gather into SIMD register};

		\node[acc, minimum width=3.5cm, minimum height=0.7cm] (fma) at (1.75,-3.7)
		{
			\normalsize $\sum_i a_i \cdot x_{c_i}$
		};
		\draw[->, thick] (simd.south) -- (fma.north);

		\node[vec] (x0) at (6.0, 1.1) {$x_0$};
		\node[vec] (x1) at (6.0, 0.55) {$x_1$};
		\node[vec] (x2) at (6.0, 0.0) {$x_2$};
		\node[vec] (x3) at (6.0,-0.55) {$x_3$};
		\node[vec, draw=none, fill=none, minimum height=0.3cm] (xdot1) at (6.0,-1.0) {$\vdots$};
		\node[vec] (x5) at (6.0,-1.45) {$x_5$};
		\node[vec] (x7) at (6.0,-2.0) {$x_8$};
		\node[vec] (x8) at (6.0,-2.55) {$x_9$};
		\node[vec, draw=none, fill=none, minimum height=0.3cm] (xdot2) at (6.0,-3.0) {$\vdots$};
		\node[vec] (x10) at (6.0,-3.55) {$x_{13}$};
		
		\node[above, font=\bfseries] at (6.0, 1.4) {Dense input vector};

		\draw[gather] (c0.east) .. controls +(0.8,0.2) and +(-0.8,0.0) .. (x2.west);
		\draw[gather] (c1.east) .. controls +(1.2,-0.2) and +(-0.8,0.0) .. (x8.west);
		\draw[gather] (c2.east) .. controls +(0.8,-0.1) and +(-0.8,0.0) .. (x5.west);
		\draw[gather] (c3.east) .. controls +(1.0,-0.5) and +(-0.8,0.0) .. (x10.west);
		\draw[gather] (c4.east) .. controls +(0.8,0.4) and +(-0.8,0.0) .. (x1.west);
		\draw[gather] (c5.east) .. controls +(0.6,-0.3) and +(-0.8,0.0) .. (x7.west);

		\node[
		draw, rounded corners, align=left,
		text width=8cm, inner sep=4pt, fill=gray!10
		] (dispatch) at (11, 0.9) {
			\textbf{Row-sweep dispatch:}
			$\ell = \texttt{row\_end} - \texttt{row\_start}$
		};
		
		\node[
		draw, rounded corners, align=left,
		text width=7cm, inner sep=4pt, fill=red!8
		] (p1) at (11,-0.2) {
			\textbf{Proc. 1:} $\ell \le 4$\\
			Empty and one-entry rows handled here.\\
			Specialized scalar microkernels.
		};
		
		\node[
		draw, rounded corners, align=left,
		text width=7cm, inner sep=4pt, fill=orange!12
		] (p2) at (11,-1.7) {
			\textbf{Proc. 2:} $5 \le \ell \le 15$\\
			Scalar unrolled dot product.\\
			Four independent partial sums.
		};
		
		\node[
		draw, rounded corners, align=left,
		text width=7cm, inner sep=4pt, fill=yellow!20
		] (p3) at (11,-3.4) {
			\textbf{Proc. 3:} $\ell \ge 16$\\
			If unit-stride input and AVX-512 ready:
			gather+FMA (2$\times$8 nonzeros/iteration).\\
			Otherwise, fall back to Proc. 2.
		};
		
		\draw[dispatcharrow] (dispatch.west) |- (p1.west);
		\draw[dispatcharrow] (dispatch.west) |- (p2.west);
		\draw[dispatcharrow] (dispatch.west) |- (p3.west);
		
	\end{tikzpicture}
	\caption{Schematic workflow of a CSR row-sweep optimization utilizing adaptive execution pathways based on row length ($\ell$). The runtime system calculates $\ell = \text{row\_end} - \text{row\_start}$ to dynamically dispatch the processing to one of three specialized procedures: Procedure 1 handles ultra-short rows ($\ell \le 4$) via scalar microkernels to eliminate loop overhead; Procedure 2 processes medium-length rows ($5 \le \ell \le 15$) using an unrolled scalar dot product targeting four independent partial sums; Procedure 3 handles long rows ($\ell \ge 16$) by leveraging hardware-specific vectorization via AVX-512 gather-and-FMA instructions when unit-stride input conditions are met, gracefully falling back to Procedure 2 otherwise.}
	\label{fig:csr_gather_kernel}
\end{figure*}
After the layout-dependent cases have been separated, the row kernel must still account for the way a CSR row reads the dense input vector. The column indices of the row select entries that are not contiguous in general, so a vector implementation needs a gather operation that loads several indexed entries into one SIMD register. This vector form is used only for unit-stride dense inputs, where the indexed addresses have the form required by the gather instruction, and only when the processor provides the AVX-512 feature set. The implementation performs the runtime feature check through \texttt{pulp}, a Rust abstraction over SIMD instructions that dispatches to vectorized code according to the detected features.\footnote{\url{https://docs.rs/pulp/latest/pulp/}} When \texttt{pulp::x86::V4::try\_new()} succeeds, the unit-stride row loop runs inside \texttt{simd.vectorize}. Empty rows, one-entry rows, and rows with at most four entries keep specialized scalar cases, and rows with five to fifteen entries use the scalar unrolled dot product. Rows with at least sixteen entries call the AVX-512 dot kernel, which processes two groups of eight nonzeros per iteration by loading column indices, gathering the corresponding dense-vector entries, loading the matrix values contiguously, and accumulating with fused multiply-add instructions; see Figure~\ref{fig:csr_gather_kernel}.

The scalar row kernel accumulates four products at a time in four independent partial sums. The column indices of the current CSR row select the corresponding entries of the dense input vector, so the dense-vector loads are indirect. The unrolled form reduces the single accumulator dependency of a scalar dot product and exposes independent product accumulations between these loads. For long unit-stride rows, the vectorized kernel processes two groups of eight nonzeros per iteration, using gather loads for the dense input entries and contiguous loads for the matrix values.

The specialized kernel uses \texttt{unsafe} only after the safe part of the routine has reduced the operation to a concrete \texttt{u32}/\texttt{f64}, one-column, sequential SpMV. At that point the remaining work is a CSR row loop whose access pattern is expressed more directly with pointers than with general safe indexing: the microkernels cast faer's generic storage to the selected representation, advance through row intervals, address the dense input through column indices and strides, and pass raw addresses to the AVX-512 gather and load intrinsics. Keeping these operations inside the inner kernel avoids repeating type, shape, and layout tests that were already performed by the dispatch, while leaving the generic public API unchanged. This use is sound because the unchecked operations are guarded by the same invariants that define valid faer matrix and vector views: the type-id checks select the concrete storage before slice casts, row offsets delimit ranges inside the index and value arrays, the dense input has the width required by the sparse matrix, and the destination pointer is derived from the single output column. The unchecked operations therefore remain inside a private kernel reached from a safe interface, implementing the low-level representation required by the microkernel while preserving the public memory-safety contract.

\subsection{Foreign backends and compilation model}\label{subsec:foreign_backends}

We expose every foreign backend to Rust through a small \texttt{extern "C"} surface for setup, execution, result extraction, and teardown. The Rust side sees only primitive arguments and an opaque backend handle, while the implementation behind that ABI boundary can use C, C++, or Fortran. Rust structs and enums do not have a C-compatible layout unless that layout is requested explicitly, so no Rust aggregate is passed by value. The compressed sparse arrays, the dense vector used by the Lanczos setups, and scalar sizes are passed as primitive pointers and integers, and setup returns the handle used by the later calls. Setup constructs the backend context, initializes dense vectors, and prepares the sparse representation accepted by the library. Eigen maps the Rust-owned CSR or CSC arrays with \texttt{Eigen::Map}, PETSc reuses the Rust-owned CSR arrays, and MKL receives the compressed arrays through its inspection-execution interface and may build internal data structures during \texttt{mkl\_sparse\_optimize}. For PSBLAS, a C++ shim initializes the context and descriptor, inserts the sparse entries with \texttt{psb\_c\_dspins}, and finalizes the internal CSR or CSC matrix with \texttt{psb\_c\_dspasb\_opt}. The PSBLAS SpMV execution calls the PSBLAS C binding directly, whereas the Lanczos executions call Fortran \texttt{bind(C)} driver routines after the same C++ shim has assembled the matrix and vectors through the PSBLAS C binding layer. Work arrays allocated by the Fortran Lanczos routines inside the execution call remain part of the PSBLAS backend measurement rather than part of the Rust harness setup.

The compilation model is chosen to avoid attributing wrapper effects to numerical kernels. The C and C++ wrappers are compiled with native optimization and link-time optimization enabled, and the Rust benchmark binary is built with fat link-time optimization. Eigen is header-only, so its sparse kernels are compiled in the wrapper translation unit under the wrapper flags. PETSc and MKL execute kernels from their installed libraries. PSBLAS SpMV uses the installed PSBLAS kernels, while the Lanczos variants run Fortran driver routines that call PSBLAS primitives. Cross-language link-time optimization applies to the Rust binary and to the C/C++ wrapper objects; the PSBLAS Fortran routines enter the final link as native objects produced by \texttt{gfortran} with \texttt{-ffat-lto-objects}.

\section{Experimental Setup and Results}\label{sec:experimental_setup}\label{sec:results}

The experiments separate the cost of the numerical kernels from setup, representation, and integration effects. We therefore first specify the execution platform, compiler stack, timed region, benchmark harness, matrix suite, and backend configurations. With these quantities fixed, we compare the native Rust implementation with PETSc, Eigen, oneMKL, and PSBLAS through performance profiles for SpMV and for one-pass and two-pass Lanczos matrix-function evaluation.

\subsection{Execution Platform and Reproducibility}

Since the comparison targets single-threaded sparse kernels, we ran all experiments on one pinned core of a quad-socket Intel Xeon Gold 6418H (\qty{60}{\mega\byte} Cache, \qty{2.10}{\giga\hertz}) server based on the Sapphire Rapids microarchitecture with \qty{2}{\tera\byte} of RAM. The system ran \texttt{Ubuntu 24.04.4 LTS} with Linux \texttt{6.8.0-111-generic} on \texttt{x86\_64}. The benchmark process was pinned to core 0 with \texttt{taskset}, library-level threading was disabled with \texttt{OMP\_NUM\_THREADS=1}, and the MKL backend was linked against \texttt{mkl\_sequential}. These choices remove scheduler migration and library-level parallelism from the measurements. Sequential execution serves as the fundamental baseline for evaluating sparse kernels, allowing us to isolate memory-safety boundaries and compiler optimizations without the interference of thread scheduling and synchronization overheads.

The foreign backends depend on a native C, C++, and Fortran software stack, so we use Spack to make that stack part of the experimental configuration. The \texttt{spack.yaml} environment uses unified concretization, GCC~14.3.0 as the required compiler, and the Sapphire Rapids target. It provides the libraries \texttt{intel-oneapi-mkl}, \texttt{eigen}, \texttt{petsc}, and \texttt{psblas}, their BLAS and MPI dependencies \texttt{openblas} and \texttt{openmpi}, and the \texttt{llvm@22} toolchain used for the C/C++ wrappers. PETSc is built in optimized shared-library mode with double precision and linked against oneMKL. PSBLAS is built from the \texttt{develop} branch with MPI support and linked against OpenMPI and OpenBLAS.

We build the Rust code with the pinned \texttt{nightly-2026-04-14} toolchain because this release uses LLVM~22, matching the Spack LLVM major version selected for the C/C++ wrappers. The benchmark profile inherits \texttt{opt-level=3}, \texttt{lto="fat"}, \texttt{codegen-units=1}, and \texttt{panic="abort"}, and benchmark runs set \texttt{RUSTFLAGS="-C target-cpu=native"}. Since the C/C++ wrapper objects participate in cross-language link-time optimization, their LLVM bitcode must be compatible with the LLVM used by \texttt{rustc}. We therefore compile them with the Spack-provided \texttt{clang}/\texttt{clang++} from LLVM~22, using \texttt{-O3 -march=native -mtune=native -flto}. The PSBLAS Fortran kernels are compiled with the Spack-provided \texttt{gfortran} and \texttt{-ffat-lto-objects}, so they provide native object code for the final link. The build script checks that the Spack LLVM installation and \texttt{rustc} have the same major version before enabling cross-language LTO.

To support reproducibility, the benchmark code and build configuration are publicly available at \href{https://github.com/lukefleed/hpla-rs}{lukefleed/hpla-rs}.

The benchmark harness exercises the backend configurations listed in Table~\ref{tab:backend_configurations}. The table separates the numerical kernel from the storage format, so the CSR and CSC entries can be interpreted as native paths or cross-format controls in the profiles below.

\begin{table}[ht]
	\centering
	\caption{Backend configurations represented in the benchmark harness.\label{tab:backend_configurations}}
	\begin{tabular}{llp{0.38\columnwidth}}
		\toprule
		\textbf{Kernel}  & \textbf{Backend} & \textbf{Configuration}           \\
		\midrule
		SpMV             & faer             & CSC, CSR                         \\
		SpMV             & Eigen            & CSC, CSR via \texttt{Eigen::Map} \\
		SpMV             & PETSc            & CSR                              \\
		SpMV             & Intel oneMKL     & CSR, CSC inspection-execution    \\
		SpMV             & PSBLAS           & CSR, CSC descriptor assembly     \\
		\midrule
		One-pass Lanczos & faer             & CSC, CSR                         \\
		One-pass Lanczos & Eigen            & CSR, CSC                         \\
		One-pass Lanczos & PETSc            & CSR                              \\
		One-pass Lanczos & PSBLAS           & CSR, CSC                         \\
		\midrule
		Two-pass Lanczos & faer             & CSC, CSR                         \\
		Two-pass Lanczos & Eigen            & CSR, CSC                         \\
		Two-pass Lanczos & PETSc            & CSR                              \\
		Two-pass Lanczos & PSBLAS           & CSR, CSC                         \\
		\bottomrule
	\end{tabular}
\end{table}

\subsection{Benchmark Design and Test Matrices}\label{subsec:methodology}\label{subsec:test_cases}

As discussed in Section~\ref{sec:sparse_matrix_kernels}, our benchmarks target three kernels: sparse matrix-vector multiplication (SpMV), one-pass Lanczos evaluation of \(\exp(-A)\mathbf{b}\), and the corresponding two-pass Lanczos scheme. We measure all three kernels with Criterion,\footnote{\url{https://docs.rs/criterion/latest/criterion/}} the Rust micro-benchmarking harness used by the benchmark crate. Criterion performs warm-up, repeated sampling, and statistical estimation for each benchmark group. In our configuration, all benchmark groups use 50 samples, a \qty{5}{\second} warm-up, and a \qty{100}{\second} measurement window. Criterion controls timing and repetition, while the harness supplies the operation count used to report throughput.

The timed region follows the setup and execution boundary defined in Section~\ref{subsec:kernel_execution_model}. For each matrix, the harness first loads the Matrix Market input, constructs the Rust-owned sparse representations, creates the backend context, and allocates the workspaces owned by the harness. Criterion then repeatedly calls only the execution routine for the selected backend. A compiler barrier is applied uniformly after each timed call, including calls through FFI, so the call result remains live. For SpMV, the timed operation is \(\mathbf{y} \leftarrow \mathbf{y} + A\mathbf{x}\), with a fixed input vector and without resetting \(\mathbf{y}\) between iterations. This avoids adding a vector initialization or copy to the kernel cost. For the Lanczos benchmarks, the timed operation includes the recurrence, the vector updates, the inner products, the projected tridiagonal computation, and the final reconstruction required by the selected one-pass or two-pass algorithm.

We normalize SpMV throughput by \(2\operatorname{nnz}\) floating-point operations per application, corresponding to one multiplication and one addition per nonzero. For one-pass Lanczos with Krylov dimension \(m\), we use \(m(2\operatorname{nnz}+11n)\), which counts the sparse matrix-vector products, the vector operations in the recurrence, and the final \(V_m \mathbf{g}_m\) accumulation. For the two-pass variant, we use \(4m(\operatorname{nnz}+4n)\), the leading-order count for the two recurrence passes and the reconstruction vector updates. These counts are used only to express throughput. The wall-clock measurements still include the complete timed execution routine.

For the SpMV benchmark, the harness iterates over all Matrix Market files present in the \texttt{matrices} directory after running \texttt{download\_matrices.sh}. The resulting 28-matrix set is listed in Table~\ref{tab:matrix_stats} and comes from the SuiteSparse Matrix Collection~\cite{SuiteSparse} (formerly known as the University of Florida Sparse Matrix Collection).
\begin{table*}[htbp]
	\centering
	\caption{SpMV test matrices from the SuiteSparse Matrix Collection~\cite{SuiteSparse}. We consider only square matrices of size $n \times n$, and report the number of nonzero entries \(\operatorname{nnz}\) after expanding symmetric Matrix Market inputs into the operator used by the benchmark harness.\label{tab:matrix_stats}}
	\begin{tabular}{llllc}
		\toprule
		\textbf{Matrix Name}   & $n$       & $\operatorname{nnz}$ & \textbf{Field/Symmetry} & \textbf{Lanczos} \\ \midrule
		Queen\_4147            & 4,147,110 & 329,499,284          & Real Symmetric          &                  \\
		amazon0302             & 262,111   & 1,234,877            & Pattern Unsymmetric     &                  \\
		as-Skitter             & 1,696,415 & 22,190,596           & Pattern Symmetric       & \faCheck         \\
		atmosmodd              & 1,270,432 & 8,814,880            & Real Unsymmetric        &                  \\
		audikw\_1              & 943,695   & 77,651,847           & Real Symmetric          &                  \\
		auto                   & 448,695   & 6,629,222            & Pattern Symmetric       & \faCheck         \\
		belgium\_osm           & 1,441,295 & 3,099,940            & Pattern Symmetric       & \faCheck         \\
		caidaRouterLevel       & 192,244   & 1,218,132            & Pattern Symmetric       & \faCheck         \\
		cant                   & 62,451    & 4,007,383            & Real Symmetric          &                  \\
		circuit5M              & 5,558,326 & 59,524,291           & Real Unsymmetric        &                  \\
		citationCiteseer       & 268,495   & 2,313,294            & Pattern Symmetric       & \faCheck         \\
		coAuthorsCiteseer      & 227,320   & 1,628,268            & Pattern Symmetric       & \faCheck         \\
		coPapersCiteseer       & 434,102   & 32,073,440           & Pattern Symmetric       & \faCheck         \\
		coPapersDBLP           & 540,486   & 30,491,458           & Pattern Symmetric       & \faCheck         \\
		consph                 & 83,334    & 6,010,480            & Real Symmetric          &                  \\
		delaunay\_n22          & 4,194,304 & 25,165,738           & Pattern Symmetric       & \faCheck         \\
		inline\_1              & 503,712   & 36,816,342           & Real Symmetric          &                  \\
		kron\_g500-logn18      & 262,144   & 21,165,908           & Integer Symmetric       & \faCheck         \\
		mac\_econ\_fwd500      & 206,500   & 1,273,389            & Real Unsymmetric        &                  \\
		pdb1HYS                & 36,417    & 4,344,765            & Real Symmetric          &                  \\
		preferentialAttachment & 100,000   & 999,970              & Pattern Symmetric       & \faCheck         \\
		rajat31                & 4,690,002 & 20,316,253           & Real Unsymmetric        &                  \\
		rgg\_n\_2\_20\_s0      & 1,048,576 & 13,783,240           & Pattern Symmetric       & \faCheck         \\
		roadNet-CA             & 1,971,281 & 5,533,214            & Pattern Symmetric       & \faCheck         \\
		shipsec1               & 140,874   & 7,813,404            & Real Symmetric          &                  \\
		smallworld             & 100,000   & 999,996              & Pattern Symmetric       & \faCheck         \\
		thermal2               & 1,228,045 & 8,580,313            & Real Symmetric          & \faCheck         \\
		web-Google             & 916,428   & 5,105,039            & Pattern Unsymmetric     &                  \\ \bottomrule
	\end{tabular}
\end{table*}
These matrices span multiple application classes, including network analysis, discretized PDE operators, and data-science workloads.

The Lanczos benchmarks use a dedicated 15-matrix symmetric subset, denoted by a \faCheck\ sign in Table~\ref{tab:matrix_stats}, because the recurrence and the a posteriori stopping rule~\eqref{eq:saad_bound} impose constraints not required by SpMV. We select matrices with small or zero mean diagonal, so the Saad a posteriori estimator for \(\exp(-A)\mathbf{b}\) is meaningful at the target tolerance. For each Lanczos matrix, we use the same deterministic starting vector for all backends. We estimate the spectral radius with a 20-step Lanczos probe, scale the matrix by the estimate before backend setup, and choose \(m\) with the Saad estimator using tolerance \(10^{-10}\), margin 50, and hard limit 500. Once \(m\) has been selected, the same value is used for every backend on that matrix.

\subsection{Analysis Framework and Performance Profiles}

Each benchmark produces one throughput value for each backend on each matrix. Since the matrices differ in size, sparsity structure, and memory-access pattern, we do not aggregate the absolute throughput values directly. We first compare the backends within each matrix, using the best measured backend on that matrix as the reference value, and then summarize these relative ratios over the whole test set. This gives each matrix one contribution to the comparison, independently of its absolute throughput scale.

Performance profiles provide this relative summary. Their standard definition uses a performance measure to be minimized, whereas our measurements report throughput, for which larger values are better. For a matrix \(i\) and a backend \(k\), if \(T_{i,k}\) denotes the measured throughput, we therefore set \(p_{i,k}=1/T_{i,k}\). The performance ratio is then \(p_{i,k}/\min_j p_{i,j}=\max_j T_{i,j}/T_{i,k}\), so a ratio equal to one identifies the backend with the largest throughput on that matrix.

A \emph{performance profile} is the collection of functions $\{ \tau_k(\alpha) : \alpha \in [1, \infty) \}$,
where $\tau_k(\alpha)$ denotes the proportion of problems for which the performance ratio of the $k$th algorithm does not exceed $\alpha$. By construction, each function $\tau_k(\alpha)$ is monotonically nondecreasing and takes values in the interval $[0,1]$.

Plotting the curves $\tau_k(\alpha)$ for multiple algorithms on the same graph shows, for each ratio \(\alpha\), the fraction of matrices on which a backend is within that factor of the best measured throughput~\cite{MR1875515}. We use this construction for the Rust implementations and the baselines listed in Table~\ref{tab:backend_configurations}.

We build separate profiles for each kernel and storage format. In the SpMV results, CSR and CSC configurations are compared in distinct profiles. In the Lanczos results, one-pass and two-pass algorithms are also kept separate. Within each profile, all curves are computed on the same set of matrices, so the comparison isolates backend performance from differences in format, algorithm, or missing measurements.

\subsection{Single-Threaded SpMV Benchmark}\label{sec:spmv}

\begin{figure*}[htbp]
	\centering
	\subfloat[CSR\label{fig:spmv_perfprof_csr}]{\includegraphics[width=\columnwidth]{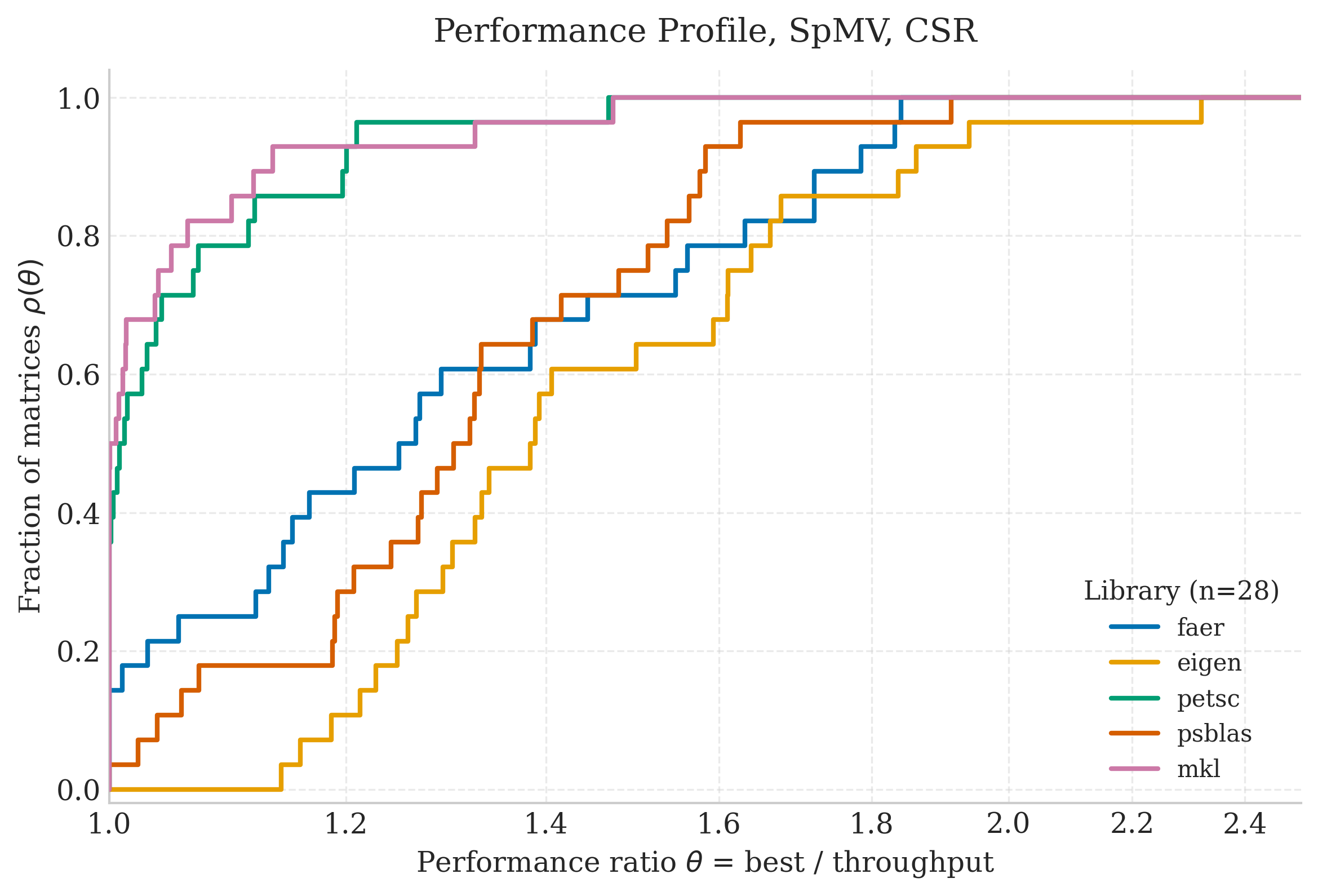}}\hfil
	\subfloat[CSC\label{fig:spmv_perfprof_csc}]{\includegraphics[width=\columnwidth]{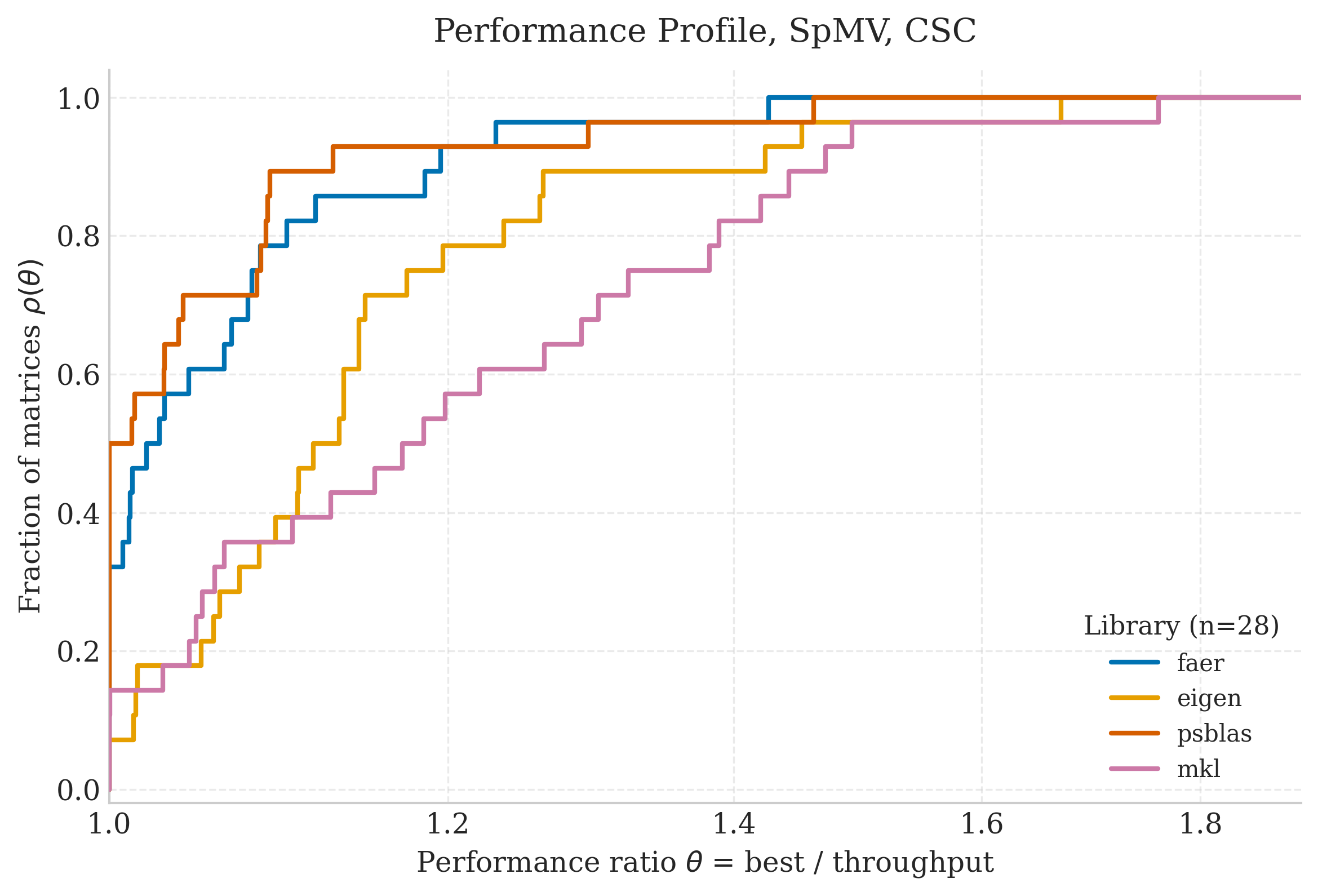}}
	\caption{Performance profiles for the single-threaded SpMV benchmark, separated by sparse storage format.}
	\label{fig:spmv_perfprof}
\end{figure*}

The SpMV experiment measures the steady-state application of an already constructed sparse operator, after matrix construction and backend setup have been completed. CSR and CSC traverse the sparse data in different orders and select different backend paths, so Figure~\ref{fig:spmv_perfprof} reports separate performance profiles for the two storage formats.

In the CSR profile, MKL attains the largest fraction of best results at ratios close to one, while PETSc remains close to the best throughput on almost the whole matrix set. The direct faer CSR kernel introduced in Section~\ref{subsec:csr_spmv_kernel} forms the next group: its curve lies below MKL and PETSc, and remains above Eigen and PSBLAS through the central part of the profile. The Rust row kernel therefore trails the strongest CSR paths over the full suite while staying within the competitive range of the single-threaded CSR implementations.

The CSC profile changes the ordering because the benchmark exercises different compressed-column paths. The faer and PSBLAS curves stay closest to the best throughput across most of the matrix set, and Eigen follows the same group over much of the plotted range. MKL, which defines one of the strongest CSR paths, no longer occupies the leading group in the compressed-column profile. The change from the CSR profile shows that a single library ranking would hide the effect of storage format and traversal order.

\begin{figure}[htbp]
	\centering
	\includegraphics[width=\columnwidth]{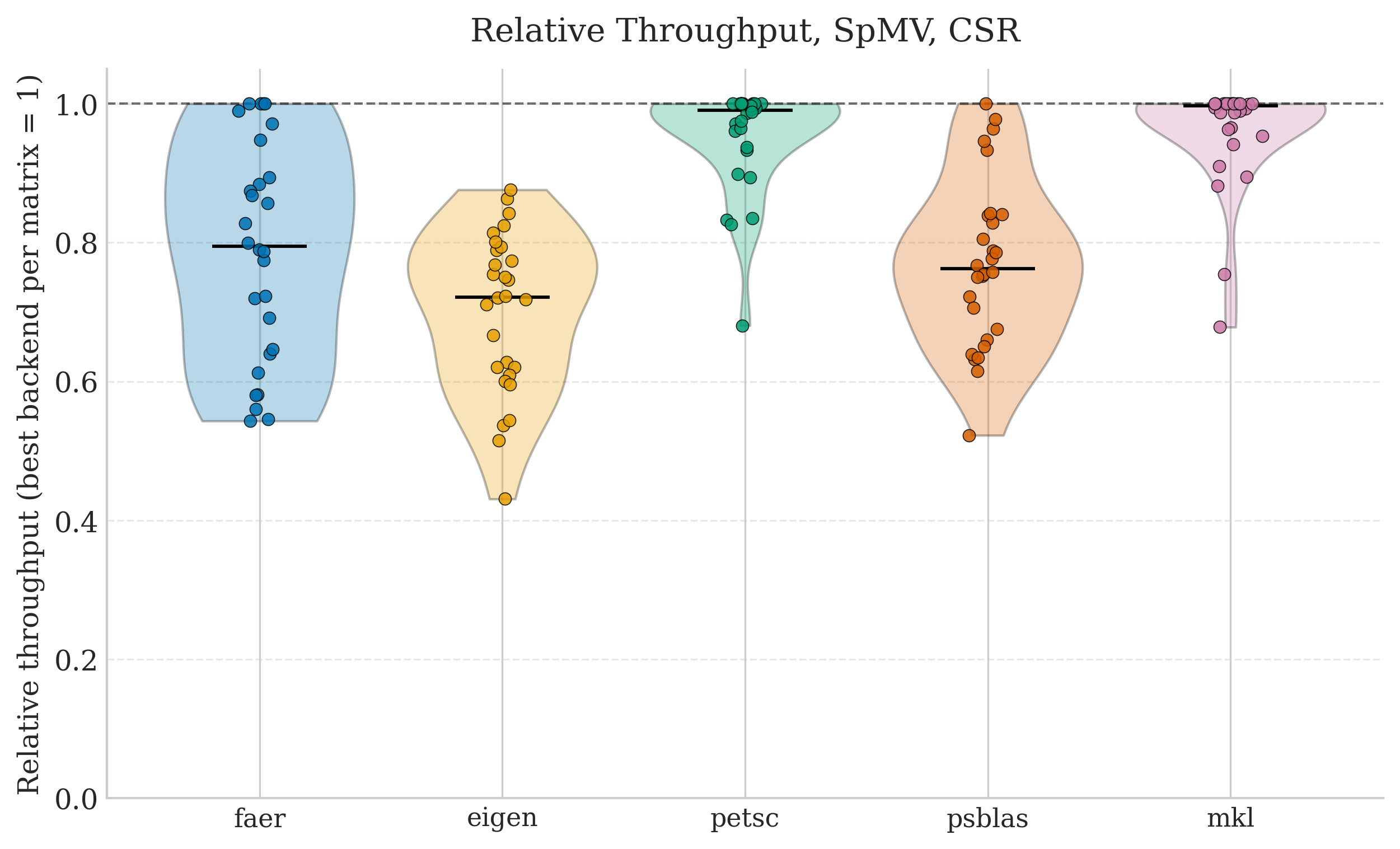}
	\caption{Distribution of relative throughput in the single-threaded CSR SpMV benchmark. For matrix $i$ and backend $k$, the plotted quantity is $T_{i,k}/\max_j T_{i,j}$, so larger values are better and the best backend on each matrix attains value $1$.}
	\label{fig:spmv_violin_csr}
\end{figure}

To complement the performance profiles, Figure~\ref{fig:spmv_violin_csr} reports a violin plot for the CSR case only. For each backend and each matrix we normalize the measured throughput by the best CSR throughput attained on that matrix, so the values lie in $(0,1]$ and the best backend for a given matrix has value $1$. The width of each violin represents the empirical distribution of these normalized throughputs across the test set, while the overlaid points show the individual matrices.
The violin plot refines the CSR interpretation. PETSc and MKL concentrate most of their mass close to one, indicating that they are not only often best in the profile sense, but also consistently close to the per-matrix optimum. The faer CSR kernel occupies the next tier, with a broader spread but a substantial fraction of matrices in the 0.8--1.0 range. Eigen and PSBLAS show lower central tendency and wider lower tails, which is consistent with the gap already visible in the CSR performance profile.

In the SpMV comparison, PETSc and MKL define the strongest CSR performance in this suite, while the specialized Rust CSR kernel occupies the next tier and stays ahead of Eigen and PSBLAS across much of the profile. PETSc's dominance is primarily driven by its Inode (Index Node) optimization, which groups rows with identical sparsity patterns to perform dense block operations. This is an advanced algorithmic layout optimization rather than a language-level speed advantage, highlighting that while the Rust CSR kernel generates highly competitive machine code, the library ecosystem has yet to integrate equivalent structural block optimizations. The compressed-column comparison is tighter for the Rust implementation, since faer tracks the best CSC throughput on most matrices. The remaining spread follows the storage format, the row or column traversal, and the sparse kernel selected by each backend.

\subsection{Lanczos Matrix-Function Benchmarks}\label{sec:lanczos_results}

The Lanczos benchmarks extend the timed region beyond the sparse matrix--vector product to encompass the complete matrix-function evaluation. Each execution therefore includes the sparse matrix--vector products, the level-1 BLAS vector operations arising in the Lanczos recurrence, the inner products and normalization steps, the projected tridiagonal matrix computation, and the final reconstruction of the approximation.

\begin{figure*}[htbp]
	\centering
	\subfloat[One-pass]{\includegraphics[width=\columnwidth]{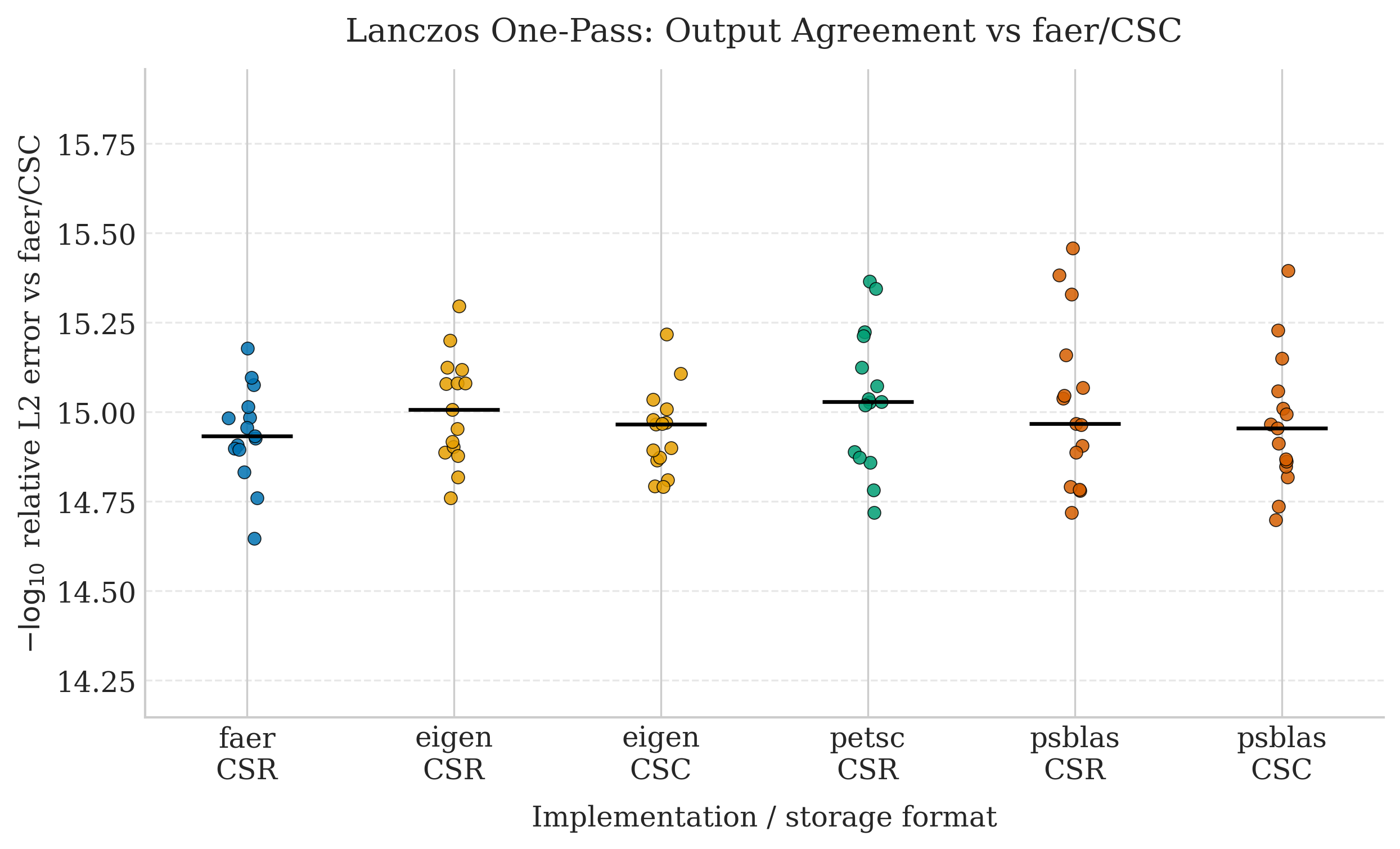}}\hfill
	\subfloat[Two-pass]{\includegraphics[width=\columnwidth]{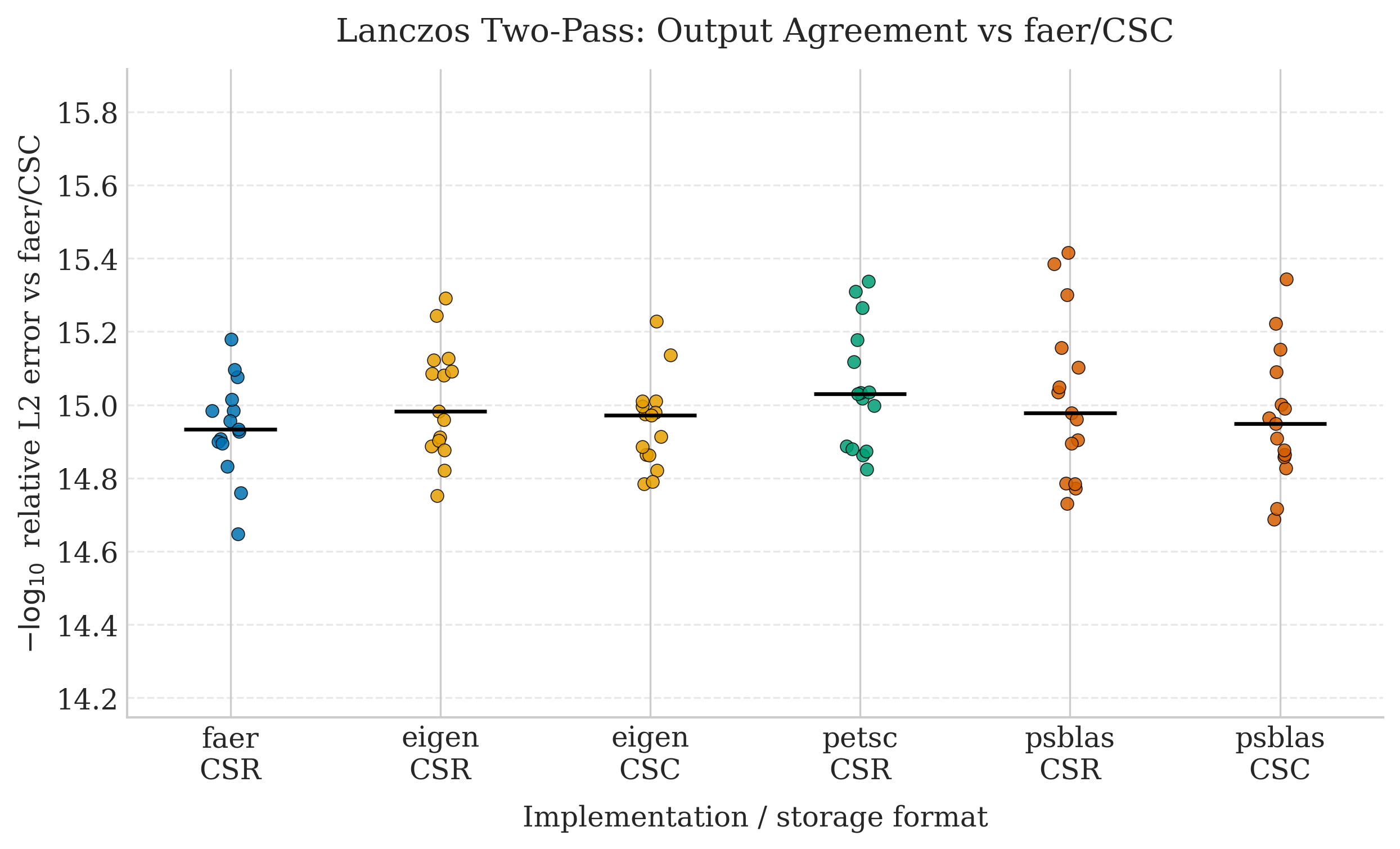}}
	\caption{Accuracy comparison (log$_{10}$ scale) of the one-pass and two-pass Lanczos implementations relative to the FAER/CSC reference implementation. Each dot represents a test matrix, and the black line denotes the mean error over all matrices.}
	\label{fig:error_comparison}
\end{figure*}

To validate the different implementations, we compare every computed vector against the FAER/CSC reference implementation and plot the relative $L_2$ difference, $\|y_k-y_{\mathrm{ref}}\|_2/\|y_{\mathrm{ref}}\|_2$, for the one-pass and two-pass variants separately. The data in Figure~\ref{fig:error_comparison} show that the non-reference implementations differ from FAER by only $3.5\times 10^{-16}$ to $2.3\times 10^{-15}$, with a median around $1.1\times 10^{-15}$ across both kernels. The same runs also record the indicator~\eqref{eq:saad_bound} used to choose the Krylov dimension; on this subset it ranges from $3.5\times 10^{-12}$ to $4.9\times 10^{-11}$, which suggests that the stopping rule is active but not close to saturation.
The results show agreement within machine precision relative to the reference across all implementations: the differences are rounding-level variations between backends, not algorithmic discrepancies. The indicator~\eqref{eq:saad_bound} remains below the prescribed tolerance throughout, so the selected Krylov dimension is sufficient for this test set.

\begin{figure*}[htbp]
	\centering
	\subfloat[One-pass, CSR\label{fig:lanczos_one_pass_csr}]{%
		\includegraphics[width=0.48\textwidth]{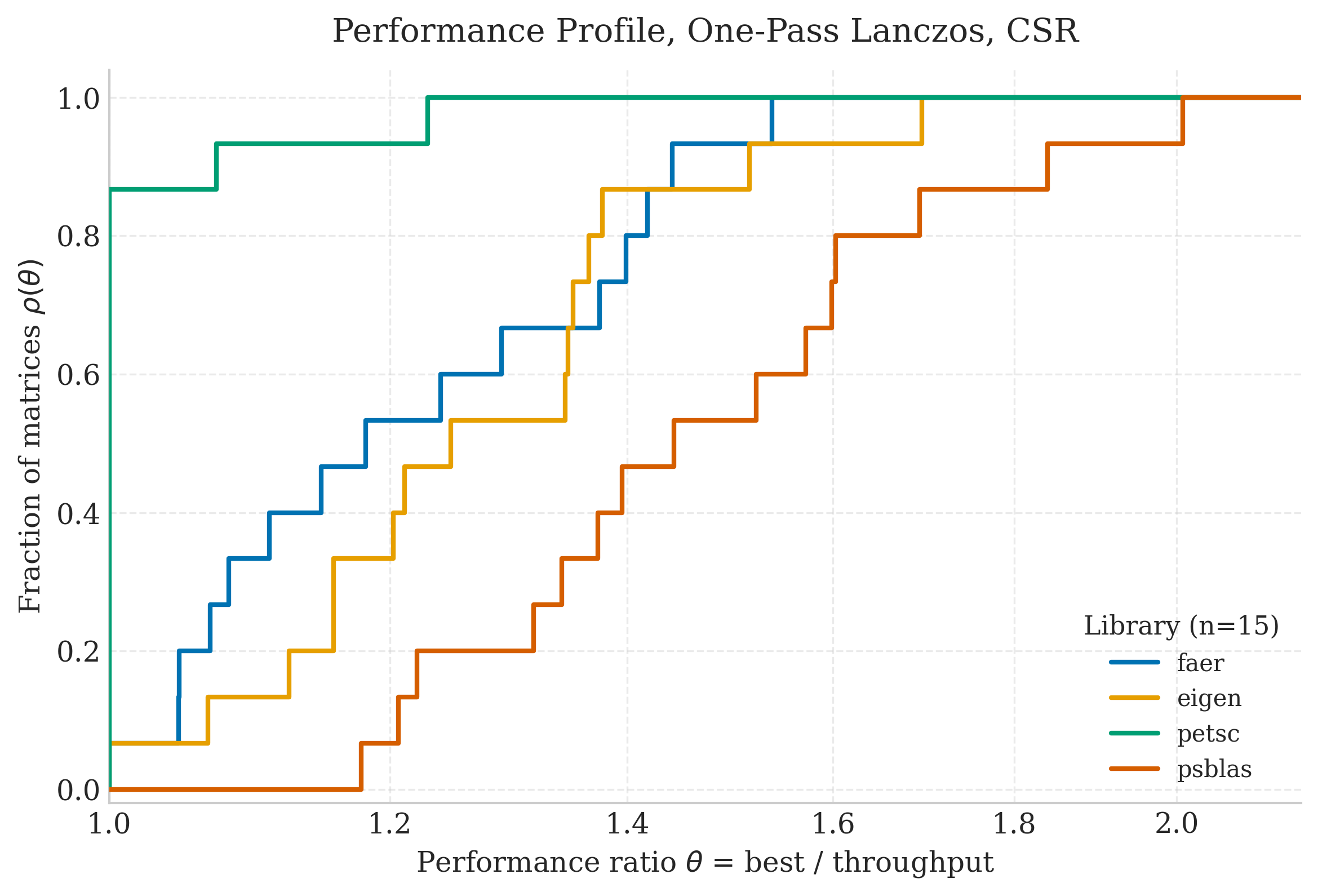}%
	}\hfil
	\subfloat[One-pass, CSC\label{fig:lanczos_one_pass_csc}]{%
		\includegraphics[width=0.48\textwidth]{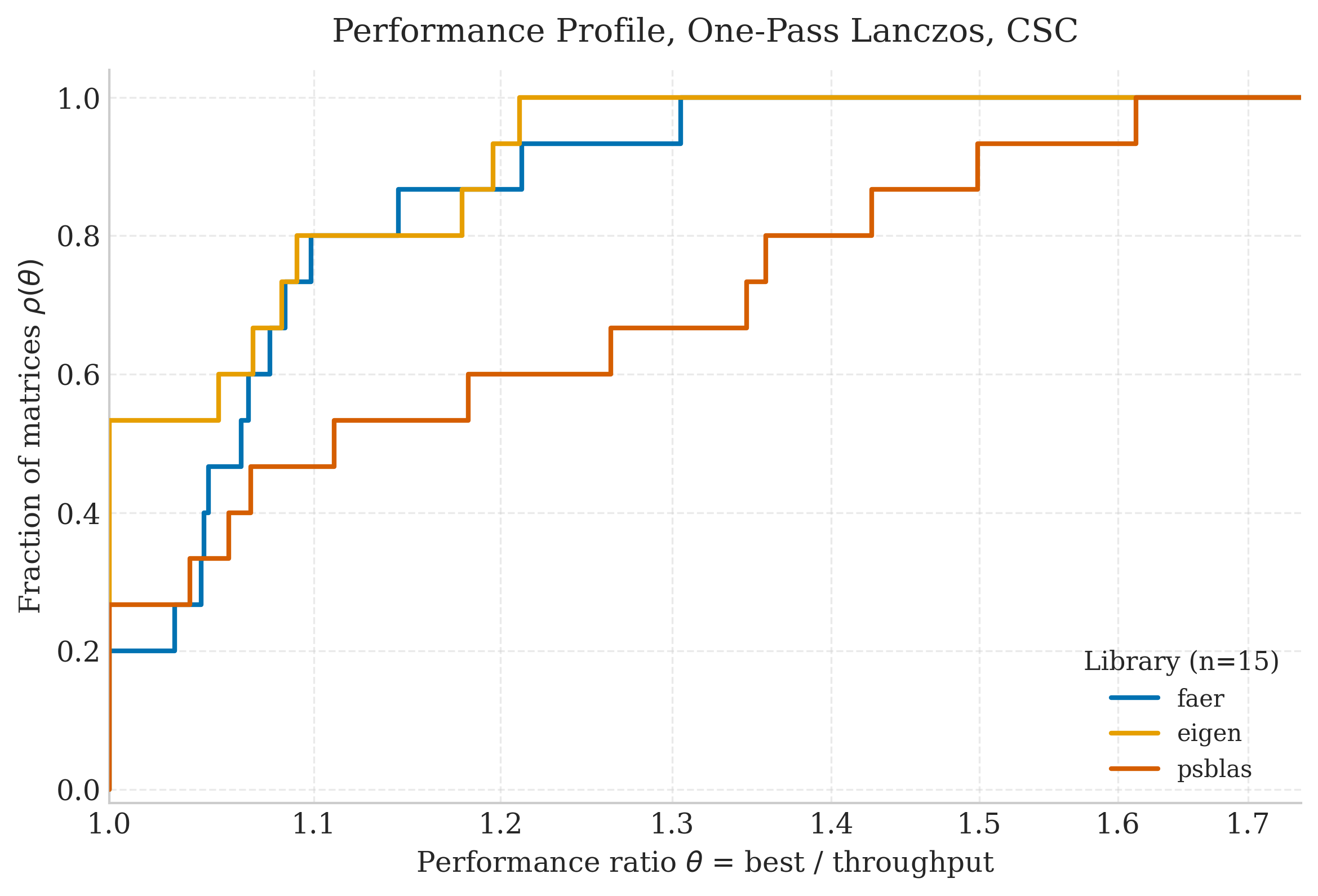}%
	}
	
	\vspace{1ex} %

	\subfloat[Two-pass, CSR\label{fig:lanczos_two_pass_csr}]{%
		\includegraphics[width=0.48\textwidth]{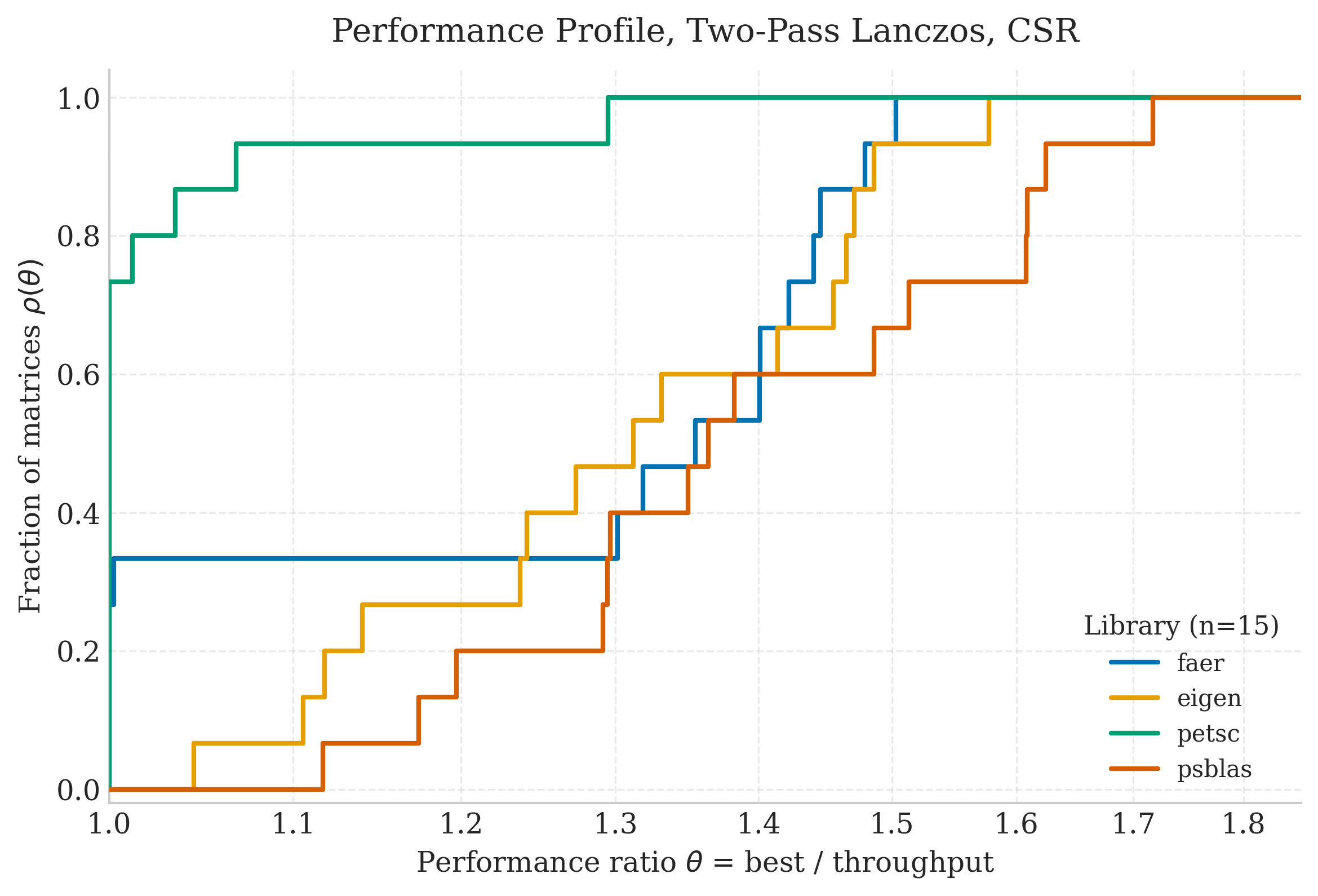}%
	}\hfil
	\subfloat[Two-pass, CSC\label{fig:lanczos_two_pass_csc}]{%
		\includegraphics[width=0.48\textwidth]{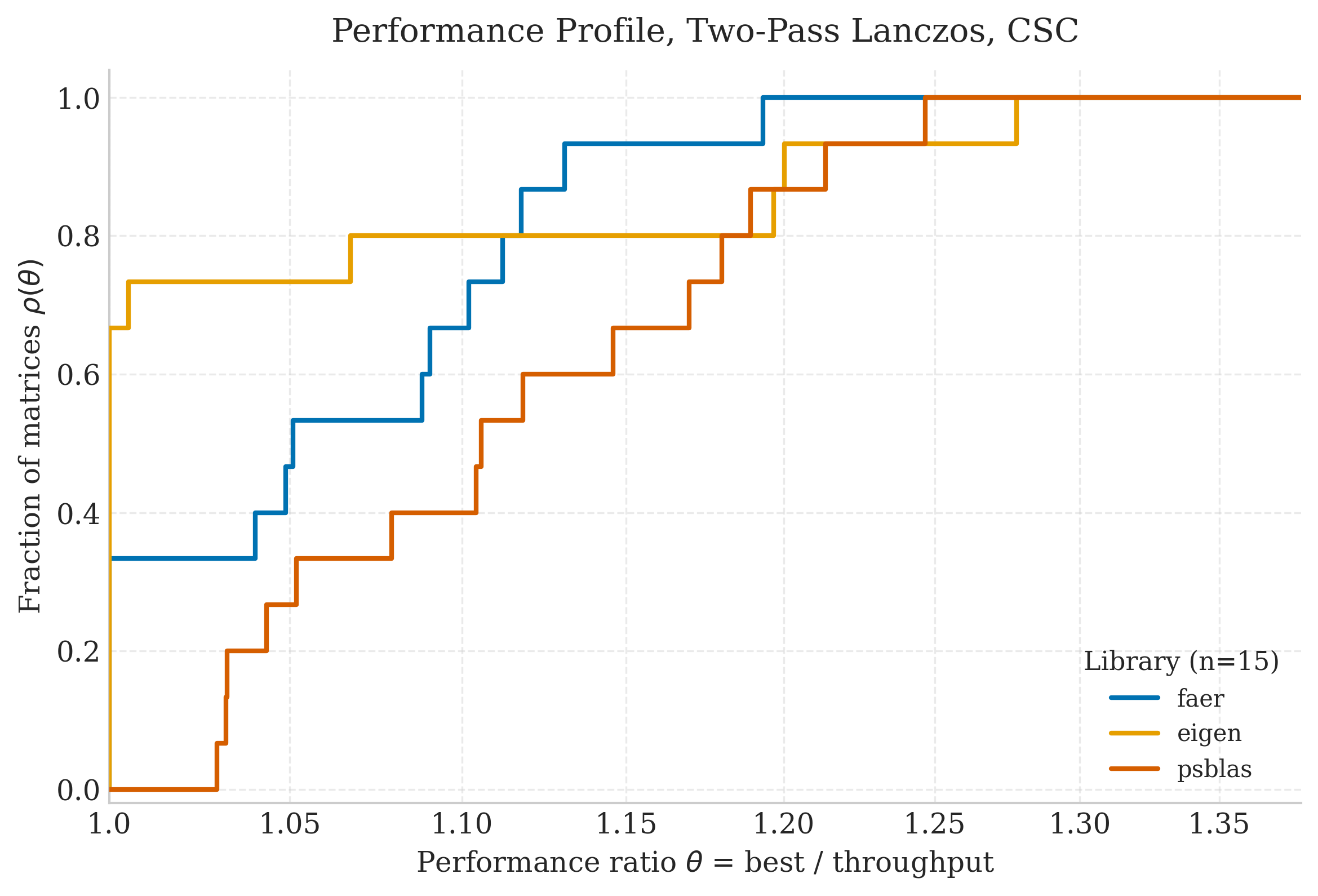}%
	}
	
	\caption{Performance profiles for the Lanczos matrix-function benchmarks, separated by algorithm and sparse storage format.}
	\label{fig:lanczos_perfprof}
\end{figure*}

Having validated the numerical consistency, we can turn to the performance of the different implementations. The profiles in Figure~\ref{fig:lanczos_perfprof} therefore measure complete one-pass and two-pass evaluations at the Krylov dimension selected for each matrix, rather than isolated SpMV calls inside the recurrence.
In the CSR profiles, PETSc defines the leading curve for both the one-pass and two-pass variants, remaining closest to the best throughput over nearly the whole Lanczos subset. Faer and Eigen form the next group. Their one-pass curves remain close, while in the two-pass case the faer curve moves slightly ahead of Eigen over the central part of the profile. PSBLAS requires larger ratios in both CSR panels. Since the timed region includes the complete recurrence and projected computation, the faer CSR curve shows that the optimized Rust recurrence preserves enough of the sparse-kernel performance to remain comparable with the other non-PETSc CSR implementations, although PETSc still dominates the CSR Lanczos profiles.

\begin{figure}[htbp]
	\centering
	\subfloat[One-pass, CSR\label{fig:violin_one_pass_csr}]{%
		\includegraphics[width=\columnwidth]{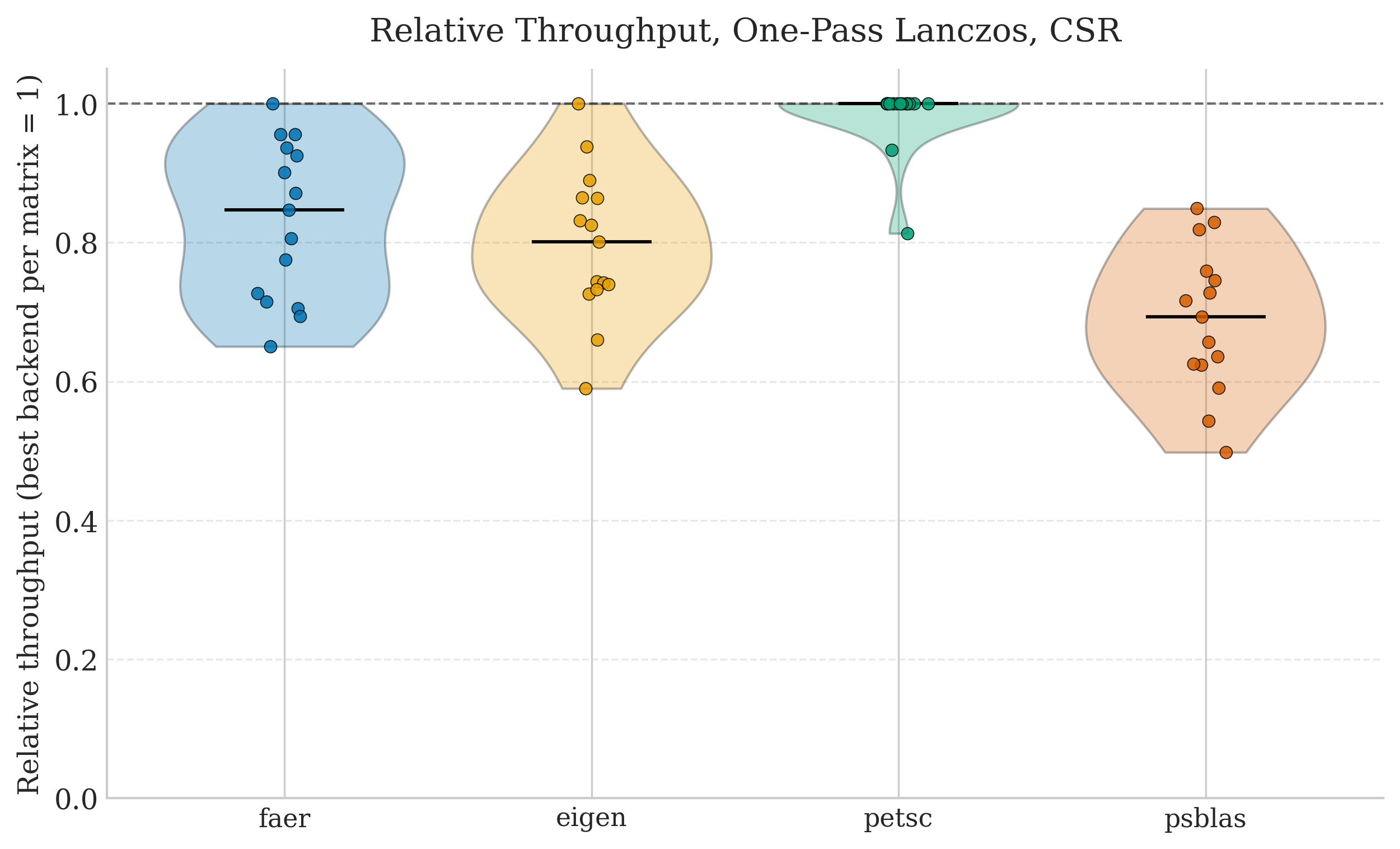}%
	}
	
	\subfloat[Two-pass, CSR\label{fig:violin_two_pass_csr}]{%
		\includegraphics[width=\columnwidth]{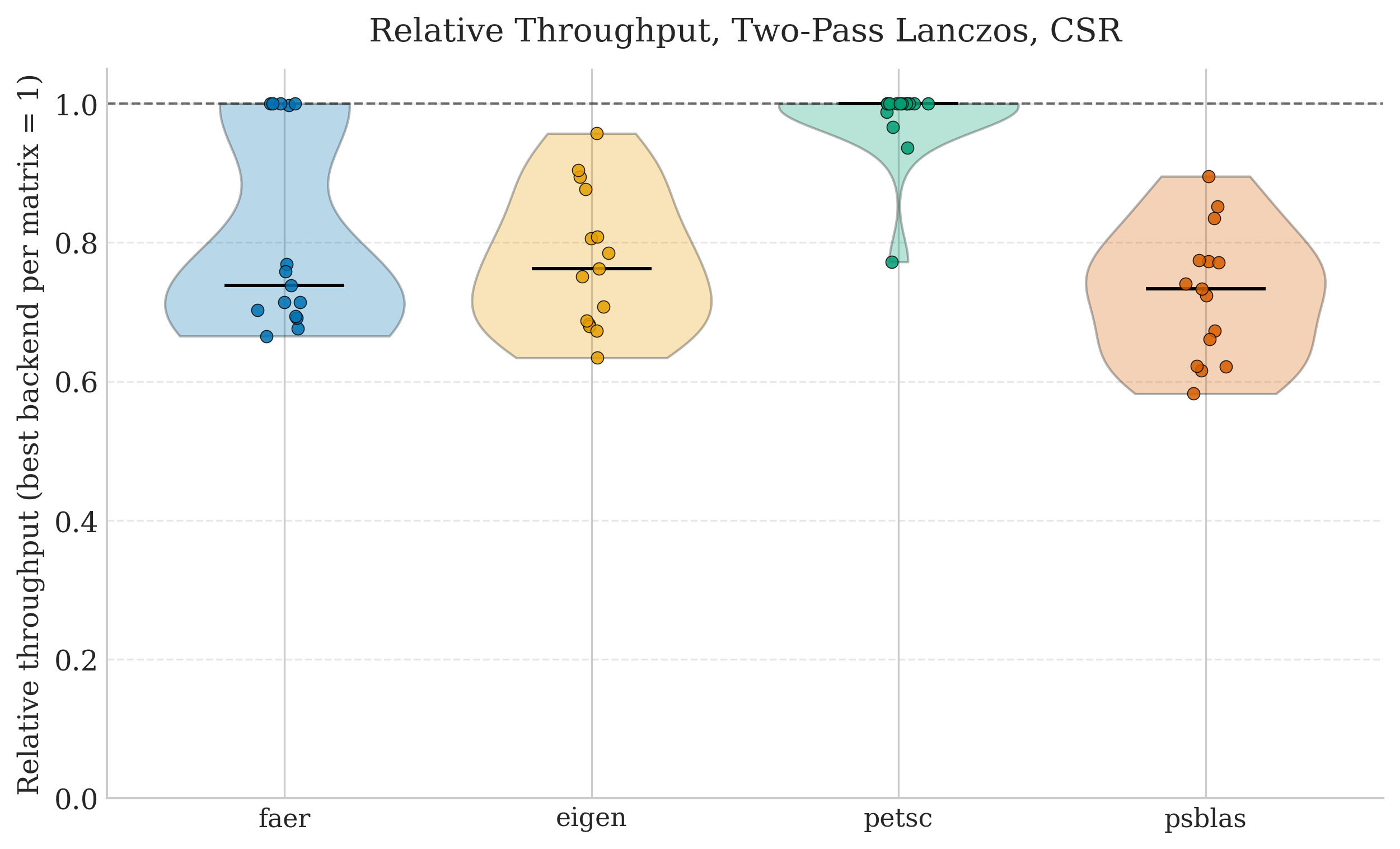}%
	}
	\caption{Distribution of relative throughput in the CSR Lanczos benchmarks. For each matrix, the best backend attains value $1$, and the remaining backends are scaled by that reference.}
	\label{fig:lanczos_violin_csr}
\end{figure}

Figure~\ref{fig:lanczos_violin_csr} gives the corresponding CSR violin plots for the normalized throughput $T_{i,k}/\max_j T_{i,j}$ in the one-pass and two-pass Lanczos experiments. Unlike the performance profiles, these plots show how stable the ranking is across matrices, with the width of each violin indicating the empirical spread and the points showing individual matrices.
These distributions confirm profile-based ranking  mainly revealing where the outliers and the spread are. PETSc is tightly concentrated near one in both variants, which indicates both the highest central tendency and the smallest degradation from the per-matrix optimum. Faer and Eigen form the intermediate group, and PSBLAS is visibly broader and shifted downward with a heavier lower tail.

The CSC profiles give the corresponding comparison for compressed-column backends. In the one-pass variant, Eigen and faer are nearly coincident over most of the profile. In the two-pass variant, faer and Eigen again form the leading group, with the faer curve occupying the front of the distribution over a substantial part of the subset. PSBLAS remains close on part of the suite, then separates at larger ratios. PETSc appears only in the CSR Lanczos comparison, so the CSC panels compare the backends that expose a compressed-column path in the harness. The change in ordering between CSR and CSC mirrors the SpMV results and shows that storage format affects both the isolated sparse product and the full matrix-function computation.

\section{Conclusions and Future Work}\label{sec:conclusions}

We evaluated Rust as an implementation language for sparse numerical kernels by comparing native Rust code with C, C++, and Fortran backends under a common benchmark harness. This comparison is demanding because the reference backends belong to language ecosystems that have defined production HPC software for decades, whereas Rust's numerical ecosystem is still comparatively young. The single-threaded SpMV profiles show that, when the storage layout and row kernel are specialized directly, a Rust CSR implementation can enter the same performance range as established sparse libraries on a broad matrix set. The Lanczos matrix-function benchmarks extend this observation to a larger computation. PETSc remains the strongest CSR backend on this test set, while the Rust implementation is comparable with Eigen and PSBLAS in the CSR profiles and belongs to the leading group in the CSC profiles, including the complete two-pass evaluation. The remaining performance gaps vary with backend and storage format rather than appearing uniformly across the Rust measurements.

The comparison also shows why Rust is relevant to high-performance numerical software beyond raw throughput. In this implementation, the benchmark harness owns the matrix storage from which every backend is initialized, foreign libraries receive borrowed views over that storage, the low-level \texttt{unsafe} operations are confined to a private CSR kernel, and monomorphization specializes the row-loop variants selected by the dispatch. Together, these choices make ownership, aliasing, layout, allocation, and specialization constraints part of the program structure while preserving native code generation and interoperability with existing C, C++, and Fortran libraries. This gives numerical code a way to record invariants that, in traditional HPC codes, are often maintained through convention, documentation, and programmer discipline. We stress also that integrating Rust into existing legacy codebases introduces FFI boundary complexities. Index offsets (such as 0-indexed C/Rust vs. 1-indexed Fortran) and structural format conversions can introduce copy overheads that must be managed carefully to preserve optimized-kernel performance.

Although the experiments in this work are deliberately single-node, production HPC codes must preserve the same ownership, layout, and interoperability constraints across process boundaries. Distributed-memory programs pass buffers to nonblocking operations, describe derived datatypes, and overlap communication with local computation; in conventional MPI codes, the validity of these actions is often maintained by call-site discipline and library documentation. Rust gives part of this structure a language-level form. Borrowing can restrict access to a buffer while a communication object depends on it, ownership can represent protocol-state transitions, and traits can associate Rust values with the datatypes used by the communication layer. The current \texttt{rsmpi} ecosystem already exposes MPI point-to-point communication, collective operations, and datatype handling to Rust programs~\cite{rsmpi}. Recent work on type-safe MPI over RSMPI develops this direction further by using typed communicators and the \texttt{Equivalence} trait to reduce datatype mismatches in point-to-point communication~\cite{iqbal2025enhancing}. Rust is also being explored beyond conventional MPI bindings, for example in distributed-memory HPC runtimes such as Lamellar~\cite{friese2024lamellar}.

The same outlook applies to heterogeneous computing. We did not benchmark accelerator kernels in this manuscript, but the literature suggests this is a plausible direction for future work. Modern sparse solvers increasingly target GPU-equipped nodes, and Rust may become more interesting for this setting when it is used to write the device kernels themselves, rather than only to launch kernels written in another language. Device code already has a restricted execution model: it must control allocation, memory layout, synchronization, address spaces, and raw hardware operations. These restrictions match Rust's strengths when kernels are written under \texttt{no\_std}-style constraints, with allocation-free code, typed device buffers, monomorphized generics, scoped atomics, barriers, and explicit \texttt{unsafe} boundaries around operations that cannot be checked by the compiler. Recent projects show that this direction is technically plausible. \texttt{cuda-oxide} compiles Rust \texttt{\#[kernel]} functions to CUDA PTX through a custom \texttt{rustc} backend~\cite{cuda_oxide}, Rust-CUDA targets CUDA execution through NVVM IR~\cite{rust_cuda}, and Rust-GPU uses \texttt{rustc\_codegen\_spirv} to compile Rust entry points annotated for SPIR-V execution into SPIR-V modules~\cite{rust_gpu}. Since SPIR-V can be consumed by Vulkan-oriented runtimes, Rust-GPU also provides a vendor-neutral route for Rust device code. A concrete next step for our work is to test these expectations directly, evaluating Rust not only for node-local kernels but also along the full path from distributed-memory operators to accelerator kernels.

\bmsubsection*{Author Contributions}

All authors conceptualized the study, contributed equally to the methodology, software development, data analysis, and the writing and editing of the manuscript.

\bmsubsection*{Funding} 
F.D. acknowledge the MUR Excellence Department Project awarded to the Department of Mathematics, University of Pisa, CUP I57G22000700001.

\bmsubsection*{Ethics Statement}
The authors have nothing to report.

\bmsubsection*{Consent}
Not applicable. No animals or humans were used in this study.

\bmsubsection*{Financial Disclosure}

None reported.

\bmsubsection*{Conflicts of Interest}

The authors declare no conflicts of interest.

\bmsubsection*{Data Availability Statement}
The matrices used in the testing are available from the SuiteSparse matrix collection \href{https://sparse.tamu.edu/}{sparse.tamu.edu/}. The GitHub repository \href{https://github.com/lukefleed/hpla-rs}{lukefleed/hpla-rs} contains all the raw data used to generate the figures together with the relevant Python scripts.

\bibliography{wileyNJD-Chicago}

\bmsubsection*{Supporting Information}

The code for running the benchmark is available from the GitHub repository \href{https://github.com/lukefleed/hpla-rs}{lukefleed/hpla-rs}.

\end{document}